\documentclass[%
 reprint,
 amsmath,amssymb,
 aps,
]{revtex4-2}

\usepackage{graphicx}
\usepackage{dcolumn}
\usepackage{bm}
\usepackage{theorem}
\usepackage{color}
\usepackage{multirow}
\usepackage{amssymb}
\usepackage{subfigure}
\usepackage{mathrsfs}
\usepackage{hyperref}

{\theorembodyfont{\normalfont\it}
\theoremheaderfont{\normalfont\bf}
\newtheorem{thm}{ Theorem}
\newtheorem{dfn}[thm]{ Definition}
\newtheorem{lmm}[thm]{ Lemma}

\newtheorem{crl}[thm]{ Corollary}
\newtheorem{asm}[thm]{ Assumption}
\newtheorem{prp}[thm]{ Proposition}
\newtheorem{cjt}[thm]{ Conjecture}
\newtheorem{rmk}[thm]{ Remark}}

{\theorembodyfont{\normalfont}
\theoremheaderfont{\normalfont\it}
\newtheorem{prf}{ Proof:}}

{\theorembodyfont{\normalfont}
\theoremheaderfont{\normalfont\it}
}

{\theorembodyfont{\normalfont}
\theoremheaderfont{\normalfont\it}
}

{\theorembodyfont{\normalfont}
\theoremheaderfont{\normalfont\it}
}

{\theorembodyfont{\normalfont}
\theoremheaderfont{\normalfont\it}
}

{\theorembodyfont{\normalfont}
\theoremheaderfont{\normalfont\it}
}

\newcommand{\bra}[1]{\mbox{$\langle#1|$}}

\newcommand{\ket}[1]{\mbox{$|#1\rangle$}}

\newcommand{\inpro}[2]{\langle#1|#2\rangle}
\newcommand{\outpro}[2]{\mbox{$\ket{#1}\!\bra{#2}$}}

\newcommand{\proj}[1]{\mbox{$\ket{#1}\!\bra{#1}$}}

\newcommand{\alg}[1]{\begin{align}#1\end{align}}

\newcommand{\nn}{\nonumber}

\newcommand{\mbb}[1]{{\mathbb #1}}

\newcommand{\bthm}[1]{\begin{thm}\label{thm:#1}}
\newcommand{\ethm}{\end{thm}}

\newcommand{\rThm}[1]{Theorem \ref{thm:#1}}
\newcommand{\blmm}[1]{\begin{lmm}\label{lmm:#1}}
\newcommand{\elmm}{\end{lmm}}

\newcommand{\bdfn}[1]{\begin{dfn}\label{dfn:#1}}
\newcommand{\edfn}{\end{dfn}}

\newcommand{\basm}[1]{\begin{asm}\label{asm:#1}}
\newcommand{\easm}{\end{asm}}

\newcommand{\bprp}[1]{\begin{prp}\label{prp:#1}}
\newcommand{\eprp}{\end{prp}}

\newcommand{\bcrl}[1]{\begin{crl}\label{crl:#1}}
\newcommand{\ecrl}{\end{crl}}

\newcommand{\bcjt}[1]{\begin{cjt}\label{cjt:#1}}
\newcommand{\ecjt}{\end{cjt}}

\newcommand{\brmk}[1]{\begin{rmk}\label{rmk:#1}}
\newcommand{\ermk}{\end{rmk}}

\newcommand{\bprf}{\begin{prf}}
\newcommand{\eprf}{\end{prf}}
\newcommand{\laeq}[1]{\label{eq:#1}}
\newcommand{\req}[1]{(\ref{eq:#1})}

\newcommand{\QED}{\hfill$\blacksquare$}
\newcommand{\lsec}[1]{\label{sec:#1}}

\newcommand{\rSec}[1]{Section \ref{sec:#1}}
\newcommand{\lapp}[1]{\label{app:#1}}

\newcommand{\rApp}[1]{Appendix \ref{app:#1}}

\newcommand{\bitem}{\begin{itemize}}
\newcommand{\entem}{\end{itemize}}
\newcommand{\benum}{\begin{enumerate}}
\newcommand{\ennum}{\end{enumerate}}

\newcommand{\otm}{\otimes}

\newcommand{\rFig}[1]{Figure \ref{fig:#1}}

\begin{document}

\preprint{APS/123-QED}

\title{Detectability of post-Newtonian classical and quantum gravity\\via quantum clock interferometry}

\author{Eyuri Wakakuwa}
 \email{e.wakakuwa@gmail.com}
\affiliation{%
Nagoya University
}%

\date{\today}

\begin{abstract}
Understanding physical phenomena at the intersection of quantum mechanics and general relativity remains a major challenge in modern physics. While various experimental approaches have been proposed to probe quantum systems in curved spacetime, most focus on the Newtonian regime, leaving post-Newtonian effects such as frame dragging largely unexplored.
In this study, we propose and theoretically analyze an experimental scheme to investigate how post-Newtonian gravity affects quantum systems. We consider two setups: (i) a quantum clock interferometry setup designed to detect the gravitational field of a rotating mass, and (ii) a scheme exploring whether such effects could be used to generate gravity-induced entanglement. Due to the symmetry of the configuration, the proposed setup is insensitive to Newtonian gravitational contributions but remains sensitive to the frame-dragging effect.
Furthermore, our scheme allows for testing whether the observed gravity-induced entanglement is consistent with the quantum equivalence principle.
While the predicted effects appear too small to detect with current technology, our scheme offers a starting point for future experiments probing post-Newtonian quantum gravitational effects.

\end{abstract}

\maketitle


\section{Introduction}

Understanding physical phenomena that lie at the intersection of quantum mechanics and general relativity remains one of the major challenges in modern physics. Recent advances in quantum control technologies, such as high-precision atom interferometry and the coherent control of increasingly massive quantum systems, have made it possible to investigate the behavior of more massive quantum systems through experiments in the laboratory. In this regime, theories suggest that gravitational effects on quantum systems may become observable. Given these developments, several table-top experiments have been proposed to test the effect of spacetime properties on quantum systems, as well as to probe the quantum aspects of gravity in the low-energy regime \cite{bose2025massive}. These proposals pave the way for exploration of the overlap between general relativity and quantum mechanics with near-future technologies.

However, most previous studies have focused on the Newtonian limit, where gravity is sufficiently weak, the speed of the object is much lower than the light speed, and spacetime is static. Consequently, the interplay between quantum effects and gravitational effects in the post-Newtonian regime have remained largely unexplored. Probing this regime may require accessing higher energy scales than those in the Newtonian regime, which in turn demands more advanced control of quantum systems. Nevertheless, the lower-order post-Newtonian regime might be comparatively more accessible than the extreme energy and length scales characteristic of particle physics, such as the Planck scale. Thus, exploring experimental setups within this regime is worthwhile for improving our approach to reconciling quantum mechanics with general relativity.

The concept of quantum clocks in curved spacetime has attracted increasing attention in recent years \cite{zych2011quantum,loriani2019interference,roura2020gravitational,arndt2015interference,margalit2015self,zych2016general,castro2017entanglement,smith2020quantum,paige2020classical,di2021gravitational}.
A quantum clock is a quantum particle with internal degrees of freedom that evolve in time.
This notion plays an important role at the intersection of relativity and quantum mechanics, as time in relativity is operationally defined through the use of clocks.
In the Newtonian regime, quantum clock interferometry has been proposed as a means to detect general relativistic time dilation \cite{zych2011quantum,roura2020gravitational,loriani2019interference}.
Given its sensitivity to proper-time differences, it is natural to ask whether such setups could be extended to probe the gravitational effects beyond the Newtonian limit.
While previous experimental proposals have primarily focused on the Newtonian regime, the goal of this work is to explore situations where post-Newtonian effects become relevant, with particular attention to the frame-dragging effect associated with rotating masses \cite{lense1918einfluss,thirring1918wirkung,mashhoon1984gravitational}.

\begin{figure}[t]
\includegraphics[bb={0 0 224 375}, scale=0.6]{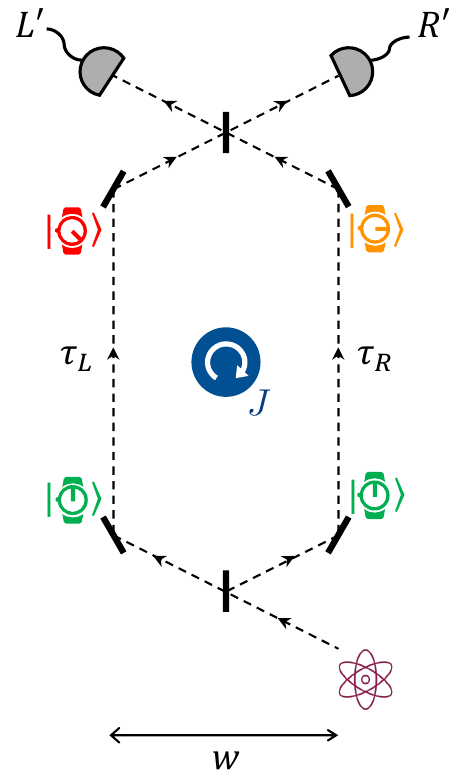}
\caption{Schematic diagram of the proposed experiment.
An atom is placed in a superposition of two parallel paths using a beam splitter and mirrors. After propagating along both paths, it is recombined to produce an interference pattern, which will be observed as a function of the width of the interferometer, $w$. A rotating massive object is located at the center of the interferometer, with its rotation axis perpendicular to the plane of the interferometer arms. While traversing each path, the atom interacts with the gravitational field generated by this object. The difference of gravitational effects between the two paths appears as a proper-time difference. Due to the symmetry of the setup, the Newtonian contributions of the gravitational field cancels out, and the remaining proper-time difference arises purely from frame-dragging. This results in the gravitomagnetic clock effect, making the observed interference pattern a direct signature of post-Newtonian gravity.}\label{fig:A}
\end{figure}

In this paper, we propose experimental setups designed to probe the effect of gravity on quantum systems, as well as to investigate the quantum nature of gravity, both in the post-Newtonian regime. Specifically, we present and analyze two experimental schemes (see \rFig{A}). The first involves testing the effect of the gravitational field generated by a rotating source mass using a quantum clock interferometer. In this setup, a quantum clock particle is split in a superposition of two parallel paths and, after levitating near the source mass, recombined to detect their interference. The gravitational interaction with the source mass induces a proper-time difference between the two paths. This leads to a modulation of the interference visibility, as originally proposed in \cite{zych2011quantum} for the case of a homogeneous gravitational field. The second experiment aims to generate gravity-induced entanglement (GIE: \cite{aspelmeyer2022zeh,marletto2025quantum,bose2025massive}) using the same interferometric scheme. Here, the source mass is prepared in a superposition of two opposite rotation directions. Since the gravitational effect on the quantum clock depends on the direction of rotation, this setup results in entanglement between the rotational degree of freedom of the source mass and both the path and internal degrees of freedom of the clock. In both experiments, the symmetry of the configuration of the interferometer ensures that Newtonian gravitational effects cancel out between the two paths. The dominant gravitational contributions in both experiments are those caused by the frame-dragging effect. Therefore, any observed change in the interference pattern can be attributed solely to post-Newtonian gravitational effects.

Furthermore, we extend the quantum equivalence principle, as formulated in Ref.~\cite{zych2018quantum}, to include frame-dragging effects and to situations in which the spacetime geometry is in a quantum superposition of classical configurations. Possible violations of this extended principle would manifest as an amplitude modulation of the interference pattern in the case of a classical gravitational field, and as a modulation of the generated GIE in the case of a superposed geometry. Accordingly, our scheme provides an experimental method capable of potentially ruling out a class of quantum-gravity models proposed to explain the mechanism by which GIE is generated.

Our quantitative analysis reveals that the frame-dragging effect on a quantum clock is exceedingly small under any realistic laboratory conditions. This suppression arises from the fact that the effect scales with the angular momentum of the rotating source and inversely with the fourth power of the speed of light. Even when employing large and fast-rotating source masses, the resulting proper-time difference remains far below the threshold of detectability.

The results highlight the limitations of tabletop experiments in accessing the quantum aspects of post-Newtonian gravity. 
 While gravitational effects in the Newtonian regime may permit experimental tests of quantum gravitational phenomena, such as entanglement generation, non-static contributions like frame-dragging appear to be experimentally inaccessible where gravitational signatures are too small to be observed.
By making this boundary explicit, our study helps delineate the range of gravitational phenomena that are accessible to current and near-future quantum experiments.

The structure of this paper is as follows.
\rSec{background} provides some background relevant to our study.
In \rSec{qCSS}, we introduce the theoretical framework describing the behavior of quantum clock particles in nonstatic but stationary curved spacetimes.
\rSec{SMRMO} evaluates the proper-time difference arising from the frame-dragging effect within this framework.
In \rSec{CIE}, we analyze the quantum clock interferometry test and its application to detecting gravity-induced entanglement.
\rSec{qEP} discusses the implications of the quantum equivalence principle in our setting.
Finally, \rSec{discussion} presents our concluding discussion.

Throughout this paper, the speed of light is denoted by $c$, the gravitational constant by $G$, and the reduced Planck constant by $\hbar$.
We adopt the signature $(-+++)$ for the spacetime metric $g_{\mu\nu}$. Greek indices such as $\mu$ and $\nu$ run over $0,1,2,3$, while Latin indices such as $i$ and $j$ run over $1,2,3$.
We also adopt the Einstein summation convention, e.g., $p_\mu q^\mu \equiv \sum_{\mu=0,1,2,3} p_\mu q^\mu$ and $p_i q^i \equiv \sum_{i=1,2,3} p_i q^i$.

\section{Background}
\lsec{background}

\subsection{Quantum Clock Interferometry}
\lsec{QCIPRL}

A quantum clock is a quantum system with an internal degrees of freedom that evolve in time and can thus serve as a reference for measuring proper time.
The concept is at the intersection of quantum mechanics and general relativity, because in general relativity, time is defined operationally through the behavior of physical clocks.
Quantum clock interferometry was proposed as a method to probe general relativistic time dilation by using quantum clocks \cite{zych2011quantum,arndt2015interference,margalit2015self,zych2016general,castro2017entanglement,loriani2019interference,smith2020quantum,roura2020gravitational,paige2020classical,di2021gravitational}. 
Unlike conventional atom interferometry \cite{roura2022quantum}, where phase shifts arise from the motion of atoms, quantum clock interferometry specifically reveals how the evolution of the internal states couples to spacetime geometry.

Theoretical frameworks describing the behavior of quantum clocks in curved spacetime have been developed in works such as \cite{zych2011quantum,pikovski2015universal,zych2017quantum,zych2018quantum,schwartz2019post,smith2020quantum}. 
These studies aim to characterize the time evolution of the internal state of a quantum clock in terms of the proper time along its trajectory, within suitable approximations.
However, formulating this behavior in a mathematically consistent way is not straightforward, even in the low-energy regime where creation and annihilation of particles are not relevant. 
This difficulty stems from the hybrid nature of the problem:
In general relativity, the proper time along a trajectory of a particle is proportional to its action integral, making the Lagrangian formalism well-suited for representing external degrees of freedom. In contrast, the internal degrees of freedom that constitute the clock often require treatment using the Hamiltonian formalism, particularly when they cannot be fully captured through standard canonical quantization.

In \cite{zych2011quantum,pikovski2015universal,zych2017quantum,zych2018quantum}, the authors derive an equation that describes the evolution of a quantum clock particle incorporating general relativistic effects. However, their result applies only to static spacetimes (where $g_{0i}=0$ for $i=1,2,3$), and therefore cannot describe systems under the frame-dragging effect. 
Another study \cite{roura2020gravitational} describes the time evolution of quantum clocks based on the assumption that the phase of each internal eigenstate follows from the classical action integral. Yet, it is not obvious whether such a formalism can be justified from the canonical quantization.

In \rSec{qCSS}, we address the above issues by developing the theoretical framework for the dynamics of quantum clock particles based on canonical quantization and the path-integral approach. In our formulation, the external degrees of freedom are described within the Lagrangian formalism, while the internal degrees of freedom are treated in the Hamiltonian formalism, in a manner similar to that of Section 2.1.4 of \cite{zych2017quantum}. As a result, the time evolution of the internal states is expressed explicitly in terms of the proper time along the trajectory of the particle.
Within the domain of validity of the approximations, this framework applies to nonstatic spacetimes and is fully justifiable from canonical quantization. It thus provides a consistent foundation for analyzing quantum clock interferometry in nonstatic spacetimes. 

Based on this framework, \rSec{DPSex} analyzes a quantum clock interferometry test designed to detect the proper-time difference induced by the frame-dragging effect. Analogously to the case of a homogeneous gravitational field\cite{zych2011quantum}, this proper-time difference manifests in the interference pattern as both a phase shift and an amplitude modulation. The present framework allows us to make quantitative predictions for these effects. We note that Ref.~\cite{basso2021effect} proposes an experimental scheme for testing the gravitomagnetic clock effect using a quantum interferometer with a spin-$1/2$ particle in place of a quantum clock.

\subsection{Gravity-Induced Entanglement}

One of the most fundamental questions in modern physics is whether gravity itself is a quantum force. 
In the realm of particle physics, it has been considered that probing the quantum nature of gravity would require extremely high-energy experiments, such as those conducted in particle accelerators or astrophysical observations of extreme environments like black holes.
However, in recent years, a novel approach has emerged, suggesting that small-scale, highly controlled tabletop experiments might provide experimental verification of quantum nature of gravity \cite{aspelmeyer2022zeh,marletto2025quantum} (see also Section IV C of \cite{bose2025massive}). In these experiments, two massive particles are placed in a spatial superposition at different locations and allowed to interact solely through gravity over a period. 
A key insight from quantum information theory states that classical interactions alone cannot generate entanglement \cite{bose2017spin} (see \cite{krisnanda2017revealing,lami2024testing,galley2022no} for general frameworks).
 Therefore, if measurable entanglement between the two particles is observed, it would serve as evidence that gravity itself must be inherently quantum in nature.
 The entanglement generated in such a setup is referred to as {\it gravity-induced entanglement}.
While there is still debate over what can be concluded from the detection of gravity-induced entanglement \cite{hall2018two,anastopoulos2021gravitational,fragkos2022inference,ma2022limits,martin2023gravity,christodoulou2023gravity,huggett2023quantum,aziz2025classical}, it nonetheless marks a significant step toward uncovering the quantum mechanical properties of gravity.

One of the most well-studied protocols is the path protocol \cite{bose2017spin,marletto2017gravitationally}, in which each of two massive spin-$1/2$ particles is placed in a superposition of two spatially separated paths. The particles interact via the Newtonian gravitational potential, and are then recombined so that the resulting path entanglement is mapped onto their spin degrees of freedom. This spin entanglement is subsequently measured to witness gravitationally  induced quantum correlations.
Another scheme is the oscillator protocol \cite{krisnanda2020observable,bose2022mechanism,yant2023gravitationally}, which involves two harmonic oscillators with delocalized wavefunctions placed in close proximity. The Newtonian gravitational interaction between them can generate entanglement directly through their continuous degrees of freedom.
An alternative proposal \cite{carney2021using} involves an atom interferometric setup, and several other related schemes have also been put forward \cite{marshman2020locality,torovs2021relative,PhysRevD.104.126030,howl2023gravitationally,fujita2023inverted,bengyat2024gravity,kaku2025sudden}.

However, most previous proposals have focused exclusively on the Newtonian regime, where all gravitational effects can be described solely by the Newtonian potential. As pointed out in \cite{chen2024quantum}, within this regime, gravity-induced entanglement can be interpreted as arising from the interaction between two source masses through a direct coupling potential, without requiring the quantization of the gravitational field as a mediator. The Newtonian potential itself is the classical solution to Einstein's equations in the weak-field and non-relativistic limit. Thus, even if experimental evidence confirms that gravity can induce entanglement, it remains possible that gravity behaves quantum mechanically in this regime while still remaining classical in the post-Newtonian regime. To date, there are very few experimental proposals that aim to demonstrate the quantum nature of gravity beyond the Newtonian regime. Notable exceptions include: Ref.~\cite{higgins2024truly}, which exploits the equivalence between mass and rotational energy to generate a superposition of gravitational fields; Ref.~\cite{lantano2024low}, which considers gravitational interaction between angular momenta of two rotating particles to produce entanglement; and Ref.~\cite{chen2024quantum}, which shows, within the framework of linearized quantum gravity, that interactions between delocalized quantum sources of gravity can give rise to entanglement that cannot be replicated using only the Newtonian potential.


In \rSec{GMEex} of this paper, we put forward a gravity-induced entanglement (GIE) experiment that has the potential to overcome the limitations discussed above. This proposed experiment goes beyond the Newtonian regime and is fundamentally post-Newtonian in nature. Specifically, it is designed in such a way that the generation of entanglement becomes sensitive to the frame-dragging effect of spacetime, which arises due to a rotating massive object. Importantly, in this setup, the Newtonian gravitational effects do not contribute to the entanglement generation, ensuring that any observed entanglement is tied to post-Newtonian effects rather than Newtonian gravity.

Another issue in previous approaches to GIE is that, even within the framework of quantum gravity theory, different models have been proposed to explain the mechanism by which gravity generates entanglement. This distinction is crucial, because the conclusions that can be drawn from observing GIE depend sensitively on which theoretical model one assumes. For example, Ref.~\cite{bose2022mechanism}, using a Fock-space description of quantized metric perturbations, argues that graviton exchange is responsible for generating GIE. In contrast, Ref.~\cite{martin2023gravity}, based on an explicit calculation of field propagators, suggests that quantum-controlled classical gravity can also give rise to GIE unless the experiment is performed over sufficiently short times. Ref.~\cite{christodoulou2023locally}, employing a path-integral formalism, claims that GIE arises from the spacetime geometry being in a quantum superposition of distinct classical configurations. A similar conclusion is reached in Ref.~\cite{chen2023quantum}, which introduces a field-basis representation of gravity.
These differences imply that, in addition to demonstrating the presence of gravity-induced entanglement, it is also essential to design experimental setups capable of discriminating, or potentially ruling out, specific quantum-gravity models underlying the generation of GIE.

In \rSec{qEP}, we propose an experimental setup that could potentially rule out one of the quantum-gravity models proposed to explain the generation of gravity-induced entanglement. The setup is based on applying the quantum equivalence principle (QEP) to tests of GIE (see also \rSec{QEPprr}). In this scheme, violations of the QEP would manifest as an amplitude modulation, both when the background spacetime is classical and when it is in a quantum superposition of classical configurations. If such a violation were observed only in the latter case, it would imply that the generation of gravity-induced entanglement cannot be explained by models in which GIE arises from a spacetime geometry in a quantum superposition of classical configurations.

\subsection{Quantum Equivalence Principle}
\lsec{QEPprr}

The Einstein equivalence principle is at the conceptual foundation of general relativity \cite{will2018theory,will2014confrontation}. It ensures that gravity can be described as a geometric property of spacetime. 
The Einstein equivalence principle comprises several subprinciples, such as the universality of free fall, the universality of gravitational redshift, and local Lorentz invariance. 
Within the domain of classical physics, these properties have been experimentally confirmed \cite{will2018theory,will2014confrontation}.

However, in the context of quantum mechanics, it is not self-evident whether Einstein's equivalence principle holds. In fact, it is not even clear how the equivalence principle should be formulated within the quantum domain.

Quantum particles can exist in superpositions of different internal energy eigenstates. It is therefore natural to consider extending the equivalence principle not only to energy eigenstates but also to such superpositions. Based on this idea, a quantum formulation of the equivalence principle in classical gravitational fields was proposed by Zych et al. in \cite{zych2018quantum}. In this framework, the particle's rest mass energy, inertial mass, and gravitational mass are each represented by corresponding internal Hamiltonian operators. The equivalence principle is expressed as the condition that these operators are identical. Experimental tests based on this model have also been reported \cite{rosi2017quantum}, examining whether the equivalence principle holds in quantum theory. The result suggests that the equivalence principle holds to a high degree of precision even in the quantum regime,  indicating that the equivalence principle, and thus the geometric description of gravity, remains valid in this domain. (Note, however, that there is some criticism in Ref.~\cite{schwartz2019post}; see Sec.~VI therein.)

When the background gravitational field is quantum, the formulation of Einstein's equivalence principle becomes an even more subtle issue. Recently, a formulation of the equivalence principle that does not rely on a specific model of quantum gravity has been proposed in \cite{giacomini2020einstein,giacomini2022quantum}. In this approach, the equivalence principle is formulated using the concept of a quantum reference frames \cite{giacomini2021spacetime,giacomini2019quantum}, assuming that the background gravitational field can be expressed as a superposition of classical gravitational fields. 
Even if the background gravitational field is quantum in nature, the requirement that some version of the equivalence principle holds may still serve as a guiding principle, ensuring that gravity can be understood as a geometric property of spacetime.

In this paper, we  pursue a different approach and adopt the phenomenological model proposed in \cite{zych2018quantum}. We begin by extending their formulation of the equivalence principle for quantum particles in classical gravitational fields to non-static spacetimes, and show that possible violations of the principle can be tested using quantum clock interferometry.
We then further extend the model to the case where the background spacetime geometry is in a quantum superposition of  classical configurations, following the assumptions in \cite{giacomini2020einstein}. Within this framework, we demonstrate that the amount of entanglement generated in quantum interferometry depends on whether the equivalence principle holds. 
Thus, it is possible in our scheme to test whether the observed gravity-induced entanglement is consistent with the equivalence principle.
 Importantly, this also provides an experimental method capable of potentially ruling out one class of quantum-gravity models proposed to explain the generation of GIE, namely, those in which entanglement is mediated by a spacetime geometry in a quantum superposition of classical configurations.

We note that an experimental proposal to detect violations of the quantum equivalence principle, based on a gravity-induced entanglement setup, was put forward in \cite{bose2023entanglement}.
We leave it for future work to investigate whether a more modern formulation of the quantum equivalence principle \cite{giacomini2020einstein,giacomini2022quantum} can be applied to our setup.

\section{Quantum Clocks in Stationary but Nonstatic Spacetime}
\lsec{qCSS}

In this section, we develop a framework for describing quantum clocks in classical curved spacetime, extending the standard treatments for static metrics to the general stationary (but nonstatic) case.
We focus on a regime where the following assumptions are satisfied:
\begin{enumerate}
\renewcommand{\theenumi}{\roman{enumi}}
\renewcommand{\labelenumi}{(\theenumi)}
\item {\it Weak gravity}: Gravity is sufficiently weak so that deviations of the spacetime metric from the Minkowski metric can be treated as a small perturbation.
\item {\it Stationary spacetime}: The spacetime metric is invariant in the coordinate time.
\item {\it Slow-motion}: The velocity of the particle is sufficiently small compared to the speed of light.
\item {\it Low energy}: The energy gap of the internal (rest) Hamiltonian is sufficiently small compared to the rest mass (i.e.~the energy of the ground state).
\end{enumerate}
Under these assumptions, we show that the time evolution of a quantum clock particle is described by a propagator as
\alg{
&\inpro{\xi_{\rm fin},q_{\rm fin};t_{\rm fin}}{\xi_{\rm ini},q_{\rm ini};t_{\rm ini}}
\nn\\
&=
\bra{\xi_{\rm fin}}\int\mathscr{D}q\sqrt{\gamma(q)}\exp{\left(\frac{\hat{H}_{rest}}{i\hbar}\int_{\rm ini}^{\rm fin}d\tau\right)}\ket{\xi_{\rm ini}}.
\laeq{PathInt}
}
Here, $(q_{\rm ini};t_{\rm ini})$ and $(q_{\rm fin};t_{\rm fin})$ are spacetime points, and $\xi_{\rm ini}$ and $\xi_{\rm fin}$ are internal states of the particle.
The integral $\int\mathscr{D}q$ is taken over all timelike trajectories from $(q_{\rm ini},t_{\rm ini})$ to $(q_{\rm fin},t_{\rm fin})$, with $\gamma$ being the determinant of the spatial components of the metric.
$\hat{H}_{rest}$ is the rest Hamiltonian of the particle, and $\int_{\rm ini}^{\rm fin}d\tau$ is the proper time along each path.
The inclusion of the metric factor $\sqrt{\gamma(q)}$ ensures invariance of the propagator under spatial coordinate transformations.

In a semiclassical limit in which the path can be regarded as a classical trajectory $P$, the propagator \req{PathInt} becomes
\alg{
&\inpro{\xi_{\rm fin},{\bm q}_{\rm fin};t_{\rm fin}}{\xi_{\rm ini},{\bm q}_{\rm ini};t_{\rm ini}}\nn\\
&=\bra{\xi_{\rm fin}}\exp{\left(\frac{\hat{H}_{rest}}{i\hbar}\int_{\rm ini}^{\rm fin}d\tau\right)}\ket{\xi_{\rm ini}}.
}
The time evolution of the internal state of the particle along a classical trajectory is thus represented by a unitary operator 
\alg{
U(P)=\exp{\left(\frac{\hat{H}_{rest}}{i\hbar}\int_Pd\tau\right)},
\laeq{UTexp}
}
confirming that the accumulated phase depends only on the proper time experienced by the particle.
This expression is consistent with the formalism obtained in \cite{zych2011quantum,pikovski2015universal,zych2017quantum,zych2018quantum} for the case of static spacetime, and will be used throughout the remaining sections of this paper.

It is instructive to note that 
\alg{
\hat{H}_{rest}\int_Pd\tau
=
\int_P\hat{H}_{rest}\frac{d\tau}{dt}dt
=
\int_P\hat{R}dt,
\laeq{routhian}
}
where $\hat{R}=\hat{H}_{rest}\frac{d\tau}{dt}$ is the Routhian (see e.g.~Section 41 in \cite{landau1982mechanics}) in which the internal degree of freedom is represented by the Hamiltonian while the external degree of freedom is represented by the Lagrangian (see Section 2.1.4 in \cite{zych2017quantum}).
We will exploit this fact later in \rSec{qEP} to formulate a quantum equivalence principle in nonstatic curved spacetime.

In the rest of this section, we provide an outline of the derivation of the propagator \req{PathInt}.
A detailed proof will be provided in \rApp{pathint}.
We follow the standard procedure to obtain a path-integral formula from the canonical quantization (see e.g.~Chapter 9.1 of \cite{peskin2018introduction} and Section 6 of \cite{srednicki2007quantum}), except that the rest Hamiltonian is kept unintegrated when transforming the Hamiltonian representation of the action integral to the Lagrangian representation.

Consider a classical particle with the rest mass $m$. For the moment, we assume that it has no internal degree of freedom.
We start with deriving an approximate classical Hamiltonian of the particle, based on the relativistic dissipation relation
\alg{
g^{\mu\nu}p_\mu p_\nu=-m^2c^2,
}
where $p_\mu$ is the four-momentum of the particle.
This equation can be solved as a quadratic equation with respect to $p_0(=E/c)$, with $E$ being the total energy of the particle. Noting that $g^{00}<0$, the positive-energy solution is given by
\alg{
\frac{E}{c}
=
-\frac{g^{0i}p_i}{g^{00}}+\sqrt{\left(\frac{g^{0i}p_i}{g^{00}}\right)^2-\frac{p_ip^i+m^2c^2}{g^{00}}}.
\laeq{Etotc}
}

Let $-2\Phi/c^2:=1+g_{00}$, and take the approximation that includes terms up to first order in $p_jp^j/m^2c^2$, $g_{0i}$, and in $\Phi/c^2$, based on the assumptions of weak gravity and slow motion (Conditions (i) and (iii) above).
The second term in Eq.~\req{Etotc} is approximately calculated to be
\alg{
&\sqrt{\left(\frac{g^{0i}p_i}{g^{00}}\right)^2-\frac{p_ip^i+m^2c^2}{g^{00}}}
\nn\\
&
\approx
mc\left(1+\frac{\Phi}{c^2}\right)+\frac{1}{2mc}\sum_{i=1}^3\frac{p_i^2}{g_{ii}}.
}
The approximate total Hamiltonian is thus given by
\alg{
H_{tot}
&\approx
E_{rest}\left(1+\frac{\Phi({\bm q})}{c^2}\right)
\nn\\
&\quad\quad+\sum_{i=1}^3\left(\frac{c^2p_i^2}{2E_{rest}g_{ii}({\bm q})}-\frac{cg_{0i}({\bm q})p_i}{g_{ii}({\bm q})}\right)\!,
\laeq{htotqpEE}
}
where ${\bm q}\equiv(q^1,q^2,q^3)$, and $E_{rest}=mc^2$ is the rest energy of the particle.

The classical Hamiltonian \req{htotqpEE} can be quantized by replacing the canonical variables $(p_i,q^i)$ with operators $(\hat{p}_i,\hat{q}^i)$ satisfying the canonical commutation relation $[\hat{p}_i,\hat{q}^j]=i\hbar\delta_{ij}$, and the rest energy $E$ by the rest Hamiltonian $\hat{H}_{rest}$. 
Note that $(\hat{p}_i,\hat{q}^i)$ and $\hat{H}_{rest}$ act on different Hilbert spaces, so that $[\hat{H}_{rest},\hat{p}_i]=[\hat{H}_{rest},\hat{q}_i]=0$.
The total quantum Hamiltonian is then given by
\alg{
\hat{H}_{tot}
&=
\hat{H}_{rest}\left(1+\frac{\Phi(\hat{\bm q})}{c^2}\right)
\nn\\
&\quad+\sum_{i=1}^3\left(\frac{c^2\hat{p}_i^2}{2\hat{H}_{rest}g_{ii}(\hat{\bm q})}-\frac{cg_{0i}(\hat{\bm q})\hat{p}_i}{g_{ii}(\hat{\bm q})}\right).
\laeq{hatHtot}
}
For the ordering of operators $\{\hat{p}_i\}$ and $\{\hat{q}^i\}$ in the second term, we adopt the {\it Weyl ordering}.
This ensures the hermiticity of the total Hamiltonian, and corresponds to the mid-point prescription when calculating the propagator.

As shown in \rApp{pathint} in detail, the short-time propagator of the Hamiltonian \req{hatHtot} is given by
\alg{
&\bra{\alpha', {\bm q}'}\exp\left(\frac{\hat{H}_{rest} \Delta t}{i\hbar}\right)\ket{\alpha, {\bm q}}
\nn\\
&\approx
\delta_{\alpha',\alpha }\left(\frac{m_0}{2\pi i\hbar \Delta t}\right)^{\frac{3}{2}}\sqrt{\gamma\left(\frac{{\bm q}+{\bm q}'}{2}\right)} \exp\left(\frac{E_{\alpha } \Delta \tau}{i\hbar}\right),
\laeq{STK}
}
where the eigendecomposition of the rest Hamiltonian is assumed to be $\hat{H}_{rest}=\sum_\alpha E_\alpha\proj{\alpha}$, and
 $m_0$ is the rest mass of the particle in the ground internal state.
$\Delta \tau$ is the proper time with respect to the coordinate time duration $\Delta t$, space displacement ${\bm q}'-{\bm q}$ and the metric $g_{\mu\nu}$.
The low-energy approximation (Condition (iv) above) is used here to eliminate the dependence of the prefactor on the rest Hamiltonian $\hat{H}_{rest}$.

Using Eq.~\req{STK}, the propagator from the initial state $(\alpha_{\rm ini},q_{\rm ini};t_{\rm ini})$ to the final state $(\alpha_{\rm fin},q_{\rm fin};t_{\rm fin})$ is given by
\alg{
&\inpro{\alpha_{\rm fin},{\bm q}_{\rm fin};t_{\rm fin}}{\alpha_{\rm ini},{\bm q}_{\rm ini};t_{\rm ini}}
\nn\\
&=\delta_{\alpha_{\rm fin},\alpha_{\rm ini}}\int\mathscr{D}q\sqrt{\gamma(q)}\exp\left(\frac{E_{\alpha_{\rm ini}}}{i\hbar}\int_{\rm ini}^{\rm fin}d\tau\right),
}
where $\int\mathscr{D}q\sqrt{\gamma(q)}$ should be interpreted as an abbreviation of
\alg{
\lim_{N\rightarrow\infty}\left(\frac{m_0}{2\pi i\hbar\Delta t}\right)^\frac{3N}{2}\left(\prod_{k=1}^{N-1}\int d{\bm q}_k\sqrt{\gamma(\bar{\bm q}_k)}\right).
}
It is now elementary to obtain Eq.~\req{PathInt}.
The complete derivation, including the treatment of operator ordering and the justification for neglecting higher-order corrections, is provided in \rApp{pathint}.

\section{Proper-Time Difference due to Frame Dragging Effect}
\lsec{SMRMO}

When a massive object rotates around an axis, the surrounding spacetime is affected in such a way that it appears to be ``dragged'' by the rotation. This phenomenon is known as the frame-dragging effect \cite{lense1918einfluss,thirring1918wirkung,mashhoon1984gravitational}. Frame dragging is a distinctive feature of general relativity and has no counterpart in Newtonian mechanics. In fact, within the framework of Newtonian gravity, the gravitational field around a rotating, axially symmetric body is identical to that of a non-rotating one, which is not the case in general relativity. Mathematically, the effect arises from the non-vanishing off-diagonal components $g_{0i}$ of the spacetime metric around a rotating mass. Therefore, all phenomena associated with the frame-dragging effect occur in the post-Newtonian regime.

The gravitomagnetic clock effect is an observable manifestation of frame dragging \cite{cohen1993standard,ciufolini1995gravitation}, whereby the proper time measured by an object in orbit around a rotating mass depends on the direction of motion.  The proper time is longer for co-rotating motion and shorter for counter-rotating motion.
The term ``gravitomagnetic'' reflects the analogy between gravity and electromagnetism: just as a moving charged particle experiences a magnetic field, a moving mass near a rotating body experiences a gravitomagnetic field. In the weak-field limit, the spacetime metric can be decomposed into a Newtonian (static) component and a rotational component, analogous to the scalar and vector potentials in electromagnetism, respectively.

In the following, we evaluate the proper-time difference caused by the gravitomagnetic clock effect in our setup.

We start with evaluating the deviation of proper time of a particle traveling from one spacetime point to another, caused by a small deviation on the background spacetime metric.
Fix a coordinate system and consider a pair of timelike spacetime points $(t_{\rm ini},\bm{x}_i)$ and $(t_{\rm fin},\bm{x}_f)$ represented by that coordinate.
Suppose that a particle travels from $(t_{\rm ini},\bm{x}_i)$ to $(t_{\rm fin},\bm{x}_f)$.
The path of the particle is determined by the condition that the action of the particle along the path takes the  minimum value over all possible passes, or equivalently, the condition that the proper time along the path is maximum.
Let $\bar{g}_{\mu\nu}$ be a time-invariant metric and let $\bar{P}$ be the path of the particle from $(t_{\rm ini},\bm{x}_i)$ to $(t_{\rm fin},\bm{x}_f)$ determined by that condition.
Likewise, let $g_{\mu\nu}=\bar{g}_{\mu\nu}+h_{\mu\nu}$ be another time-invariant metric and $P$ be the path of the particle from $(t_{\rm ini},\bm{x}_i)$ to $(t_{\rm fin},\bm{x}_f)$ that minimizes the proper time with respect to that metric.
The deviation of proper time between the two paths are given by
\alg{
\Delta \tau
=
\int_P\frac{d\tau}{dt}dt
-
\int_{\bar{P}}\frac{d\bar{\tau}}{dt}dt,
\laeq{DeltaS}
}
where $t$ is the coordinate time, and $\tau$ and $\bar{\tau}$ are proper times of the particle in the background spacetime metrics $g_{\mu\nu}$ and $\bar{g}_{\mu\nu}$, respectively.
Under the condition that $h_{\mu\nu}$ is a small perturbation in the sense that $|\Lambda|\ll1$, where
\alg{
\Lambda:=\frac{h_{\mu\nu}\frac{dx^\mu}{dx^0}\frac{dx^\nu}{dx^0}}{\bar{g}_{\mu\nu}\frac{dx^\mu}{dx^0}\frac{dx^\nu}{dx^0}},\laeq{hsmallpert}
}
we prove in the following that
\alg{
\Delta \tau
=
\frac{1}{2}\int_{\bar{P}}\frac{h_{\mu\nu}}{c^2}\frac{dx^\mu}{d\bar{\tau}}\frac{dx^\nu}{d\bar{\tau}}d\bar{\tau}+O(\Lambda^2)\int_{\bar{P}}(dt+d\bar{\tau}).
\laeq{devPTdeltaS}
}
In particular, when $h_{\mu\nu}$ has only $h_{0i}$ and $h_{i0}$ components ($i=1,2,3$), we have, in the slow-motion and weak-gravity approximation,
\alg{
\!\Delta \tau
=
\int_{\bar{P}}\frac{h_{0i}}{c}\left(\frac{dt}{d\bar{\tau}}\right)dx^i+T\cdot O\left(\left|h_{0i}\frac{dx^i}{dx^0}\right|^2\right),\!
\laeq{devPTdeltaSh0i}
}
where $T:=\int_{\bar{P}}dt$ is the time scale of the path.

We start with expanding $\frac{d\tau}{dt}$ in terms of $\Lambda$. We have
\alg{
\frac{d\tau}{dt}
&=
\sqrt{-g_{\mu\nu}\frac{dx^\mu}{dx_0}\frac{dx^\nu}{dx_0}}
\\
&=
\sqrt{-(\bar{g}_{\mu\nu}+h_{\mu\nu})\frac{dx^\mu}{dx^0}\frac{dx^\nu}{dx^0}}
\\
&=
\sqrt{-\bar{g}_{\mu\nu}\frac{dx^\mu}{dx^0}\frac{dx^\nu}{dx^0}}
\sqrt{1+\frac{h_{\mu\nu}\frac{dx^\mu}{dx^0}\frac{dx^\nu}{dx^0}}{\bar{g}_{\mu\nu}\frac{dx^\mu}{dx^0}\frac{dx^\nu}{dx^0}}}
}
\alg{
&=
\sqrt{-\bar{g}_{\mu\nu}\frac{dx^\mu}{dx^0}\frac{dx^\nu}{dx^0}}
\left(1+
\frac{1}{2}\frac{h_{\mu\nu}\frac{dx^\mu}{dx^0}\frac{dx^\nu}{dx^0}}{\bar{g}_{\mu\nu}\frac{dx^\mu}{dx^0}\frac{dx^\nu}{dx^0}}+O(\Lambda^2)\right)
\\
&=\frac{d\bar{\tau}}{dt}
\left(
1-
\frac{1}{2}\frac{h_{\mu\nu}}{c^2}\frac{dx^\mu}{d\bar{\tau}}\frac{dx^\nu}{d\bar{\tau}}+O(\Lambda^2)
\right),
}
where the last line follows due to $\sqrt{-\bar{g}_{\mu\nu}\frac{dx^\mu}{dx^0}\frac{dx^\nu}{dx^0}}=\frac{d\bar{\tau}}{dt}$.
Hence, the deviation of proper time due to the perturbation of the metric (but along the same path) is given by
\alg{
\delta'\!\left(\frac{d\tau}{dt}\right)
&:=
\frac{d\tau}{dt}-\frac{d\bar{\tau}}{dt}\\
&=
\frac{d\bar{\tau}}{dt}\left(-\frac{1}{2}\frac{h_{\mu\nu}}{c^2}\frac{dx^\mu}{d\bar{\tau}}\frac{dx^\nu}{d\bar{\tau}}+O(\Lambda^2)\right).
\laeq{devPT}
}

We now express \req{DeltaS} as
\alg{
\Delta \tau
=&
\left(\int_P\frac{d\bar{\tau}}{dt}dt
-
\int_{\bar{P}}\frac{d\bar{\tau}}{dt}dt\right)
\nn\\
&
\quad+
\left(\int_P\delta'\!\left(\frac{d\tau}{dt}\right)dt
-
\int_{\bar{P}}\delta'\!\left(\frac{d\tau}{dt}\right)dt\right)
\nn\\
&
\quad+
\left(\int_{\bar{P}}\frac{d\tau}{dt}dt
-
\int_{\bar{P}}\frac{d\bar{\tau}}{dt}dt\right)
\\
& =
\int_{\bar{P}}\frac{d\tau}{dt}dt
-
\int_{\bar{P}}\frac{d\bar{\tau}}{dt}dt
+O(\Lambda^2)\int_{\bar{P}}dt.
\laeq{pbarh}
}
The last line follows from the fact that $\bar{P}$ is the path for which the proper time is the maximum with respect to the metric $\bar{g}_{\mu\nu}$, and that  both $\delta'(\frac{d\tau}{dt})$ and the deviation of the paths are of order linear in $\Lambda$.
The integral in Eq.~\req{pbarh} is calculated by using \req{devPT} to be
\alg{
&\int_{\bar{P}}\left(\frac{d\tau}{dt}
-
\frac{d\bar{\tau}}{dt}\right)dt
\nn\\
&=
-\frac{1}{2}\int_{\bar{P}}\frac{h_{\mu\nu}}{c^2}\frac{dx^\mu}{d\bar{\tau}}\frac{dx^\nu}{d\bar{\tau}}d\bar{\tau}
+O(\Lambda^2)\int_{\bar{P}}d\tau,
}
hence we arrive at \req{devPTdeltaS}.
When $h_{\mu\nu}$ has only $h_{0i}$ and $h_{i0}$ components, 
\alg{
\frac{h_{\mu\nu}}{2c^2}\frac{dx^\mu}{d\bar{\tau}}\frac{dx^\nu}{d\bar{\tau}}d\bar{\tau}
=
\frac{h_{0i}}{c^2}\frac{dx^0}{d\bar{\tau}}\frac{dx^i}{d\bar{\tau}}d\bar{\tau}
=
\frac{h_{0i}}{c}\frac{dt}{d\bar{\tau}}dx^i
}
and
\alg{
h_{\mu\nu}\frac{dx^\mu}{dx^0}\frac{dx^\nu}{dx^0}
=
2h_{0i}\frac{dx^i}{dx^0}.
}
In the slow-motion and weak-gravity approximation, we have $\frac{d\bar{\tau}}{dt}=|\bar{g}_{\mu\nu}\frac{dx^\mu}{dx^0}\frac{dx^\nu}{dx^0}|=O(1)$. This implies $O(\Lambda^2)=O(|h_{0i}\frac{dx^i}{dx^0}|^2)$, and therefore \req{devPTdeltaSh0i}.

The above argument is summarized as follows.
When the initial and final points in spacetime are fixed, and a small perturbation is introduced to the spacetime metric, the deviation in proper time arises from two contributions:
(i) the deviation of the paths, and
(ii) the change in the metric itself.
The first contribution can be evaluated along the original metric, and the second along the original path, since cross-terms are of the second order and can be neglected.
Contribution (i)  is also of the second order because the original path minimizes the proper time under the unperturbed metric.
Therefore, to first order in the perturbation, the deviation in proper time is solely due to (ii).

Suppose that $h_{\mu\nu}$ has only $h_{0i}$ and $h_{i0}$ components and $\bar{g}_{0i}=\bar{g}_{i0}=0$.
Let $\bar{P}_+$ and $\bar{P}_-$ be two paths that are time reversal of each other, that is;
\alg{
&\bar{P}_+:(t_{\rm ini},\bm{x}_A)\rightarrow(t_{\rm fin},\bm{x}_B),
\nn\\
&\bar{P}_-:(t_{\rm ini},\bm{x}_B)\rightarrow(t_{\rm fin},\bm{x}_A).
\nn
}
Both paths are considered to be the one that maximize the proper time under the metric $\bar{g}_{\mu\nu}$.
Likewise, let $P_+$ and $P_-$ be the paths that have the same initial and final points as  $\bar{P}_+$ and $\bar{P}_-$, respectively,
which maximize the proper time under $g_{\mu\nu}$.
Note that the proper time of $\bar{P}_+$ and $\bar{P}_-$ under the metric $\bar{g}_{\mu\nu}$ are the same.
Using the above result, the difference of proper times between the paths $P_+$ and $P_-$ under $g_{\mu\nu}$ is evaluated as
\alg{
&\tau(P_+)-\tau(P_-)
\nn\\
&=
\Delta\tau(P_+)-\Delta\tau(P_-)
\\
&=
-\frac{2}{c}\int_{\bar{P}_+}h_{0i}\left(\frac{dt}{d\bar{\tau}}\right)dx^i+T\cdot O\left(\left|h_{0i}\frac{dx^i}{dx^0}\right|^2\right)\\
&=
-\frac{2E}{mc^2}\int_{\bar{P}_+}\frac{h_{0i}}{c\bar{g}_{00}}dx^i+T\cdot O\left(\left|h_{0i}\frac{dx^i}{dx^0}\right|^2\right),
\laeq{h0ig00}
}
where $m$ is the rest mass of the particle and $E$ is the total energy of the particle that is conserved throughout the path.
Equation \req{h0ig00} quantifies the extent to which global clock synchronization is impossible in that coordinate system (see, e.g., Section 84 of \cite{landau1975classical}).
It should be noted that the ratio $E/mc^2$ is independent of the rest mass of the particle, as expected from the equivalence principle.

\begin{figure}[t]
\includegraphics[bb={0 0 226 222}, scale=0.8]{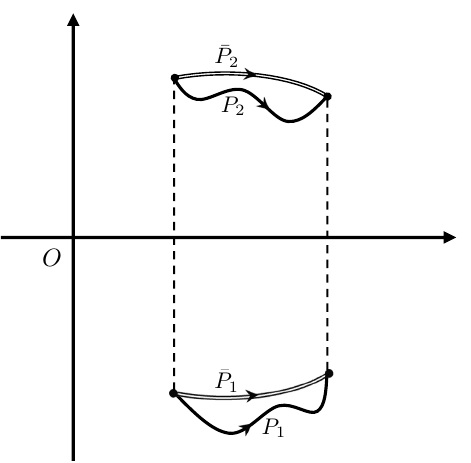}
\caption{ The paths for which the proper time difference is evaluated.}\label{fig:PTD}
\end{figure}

Now we consider the case of  an axially symmetric metric. In the polar coordinate, both $\bar{g}_{\mu\nu}$ and $h_{\mu\nu}$ depends only on the radius $r$ and the latitude $\theta$. 
Furthermore, $h_{\mu\nu}$ only has $h_{t\phi}=h_{\phi t}$ components, where $\phi$ is the longitude. 
Let $P_1$ and $\bar{P}_1$ be two paths with the same initial and final spacetime points, minimizing the proper time under the metrics $g_{\mu\nu}$ and $\bar{g}_{\mu\nu}$, respectively. Define $P_2$ and $\bar{P}_2$ similarly (see \rFig{PTD}).
Suppose that $\bar{P}_1$ and $\bar{P}_2$ are converted to each other by $\phi\rightarrow-\phi$. 
Due to the symmetry of the metric, the transformation $\phi\rightarrow-\phi$ is equivalent to $t\rightarrow-t$.
The difference of the proper time of the paths $P_1$ and $P_2$ are, in the same way as \req{h0ig00}, given by
\alg{
\tau(P_1)-\tau(P_2)
=
\frac{2E}{mc^2}\int_{\bar{P}_1}\frac{h_{t\phi}}{c\bar{g}_{tt}}d\phi+T\cdot O\left(\left|\frac{h_{t\phi}}{c}\frac{d\phi}{dt}\right|^2\right).
\laeq{htpgtt}
}

An important point to note is that equations \req{devPTdeltaS}, \req{devPTdeltaSh0i}, \req{h0ig00} and \req{htpgtt} are all scalar quantities, and are therefore independent of the particular choice of coordinate system.
This will ensure that the observed phase shift cannot be eliminated by a coordinate transformation and is a genuine manifestation of spacetime curvature.

Let us consider spacetime around a rotating axially symmetric massive object.
At the points sufficiently far from the object, the spacetime metric is, in the Boyer-Linquist coordinate, given by
\alg{
\!-(cd\tau)^2
=
&-\left(1-\frac{2GM}{c^2r}\right)(cdt)^2
+\left(1+\frac{2GM}{c^2r}\right)dr^2
\nn\\
&
\!\!\!\!\!+r^2(d\theta^2+\sin^2\!\theta d\phi^2)-\frac{4GJ}{c^3r}\sin^2\!\theta(cdt)d\phi,\!\!\!
}
where $M$ and $J$ are mass and angular momentum of the rotating object, respectively (see e.g.~Eq.~(6.1.1) in \cite{ciufolini1995gravitation}).
The first three terms are equal to the Schwalzschild metric (up to an approximation in the second term), and the last term represents the frame dragging effect caused by the rotation of the source mass.
Let $\bar{g}_{\mu\nu}$ denote the Schwarzschild part of the metric and $h_{\mu\nu}$ be the frame dragging one, that is,
\alg{
&\bar{g}_{tt}=-1+\frac{2GM}{c^2r},\;\bar{g}_{rr}=1+\frac{2GM}{c^2r},\laeq{gttgrr}\\
&\bar{g}_{\theta\theta}=r^2,\;\bar{g}_{\phi\phi}=r^2\sin^2\!\theta
}
and
\alg{
h_{t\phi}=h_{\phi t}=-\frac{2GJ}{c^3r}\sin^2\!\theta,\laeq{htp}
}
with all the other elements equal to zero.
In particular, in the equatorial plane, $\theta=\pi/2$.

We now substitute \req{gttgrr} and \req{htp} to \req{htpgtt}, and evaluate the difference of the proper time.
The integrand in the first term is calculated as
\alg{
\frac{h_{t\phi}}{c\bar{g}_{tt}}
=
\frac{1}{c}\frac{2GJ}{c^3r}\left(1+\frac{2GM}{c^2r}+o\left(\frac{GM}{c^2r}\right)\right).
\laeq{bach}
}
Noting that $r\frac{d\phi}{dt}$ is the longitudinal component of the velocity $v$ of the particle, the integral of the second term in \req{bach} is evaluated to be
\alg{
\int_{\bar{P}_1}\frac{1}{c}\frac{GM}{c^2r}\frac{GJ}{c^3r}d\phi
&=
\int_{\bar{P}_1}\frac{GM}{c^2r}\frac{GJ}{c^3r^2}\frac{1}{c}r\frac{d\phi}{dt}dt
\\
&=T\cdot O\left(\frac{GM}{c^2r}\frac{GJ}{c^3r^2}\frac{v}{c}\right).
}
The order of the second term in \req{htpgtt} can be evaluated similarly.
Thus, we have the deviation of the proper time
\alg{
\Delta\tau
&:=
\tau(P_1)-\tau(P_2)\\
&=
\frac{E}{mc^2}\frac{4GJ}{c^4}\int_{\bar{P}_1}\frac{d\phi}{r} +T\cdot O\left(\frac{GM}{c^2r}\frac{GJ}{c^3r^2}\frac{v}{c}\right).
\laeq{DeltaTauGL}
}

\section{Clock Interferometry Experiments}
\lsec{CIE}

We consider the setup of an atom interferometer depicted in \rFig{A}.
The atom has an internal degree of freedom whose dynamics is described by the rest Hamiltonian $\hat{H}_{rest}$ and serves as a quantum clock.
For the simplicity of analysis, we assume that the clock is described as a two-level system composed of the ground state $|g\rangle$ and an excited state $|e\rangle$.
The rest Hamiltonian $\hat{H}_{rest}$ is then $\hat{H}_{rest}=E_g\proj{g}+E_e\proj{e}$, where $E_g$ and $E_e$ are the energy eigenvalues of the ground state and the excited state, respectively.
Due to  Equation \req{UTexp}, the evolution of the clock state along a path $P$ is represented by the unitary operator
\alg{
U(P)&=e^{\frac{E_g\tau(P)}{i\hbar}}\proj{g}+e^{\frac{E_e\tau(P)}{i\hbar}}\proj{e}\\
&=e^{\frac{\bar{E}\tau(P)}{i\hbar}}\left(e^{-\frac{\Delta E\tau(P)}{i\hbar}}\proj{g}+e^{\frac{\Delta E\tau(P)}{i\hbar}}\proj{e}\right),
}
where $\bar{E}:=(E_g+E_e)/2$ is the average rest mass,  $\Delta E:=(E_e-E_g)/2$ is half of the energy difference, and $\tau(P)$ is the proper time along the trajectory $P$.
The state thus acquires a phase depending on the path and the internal eigenstate, which leads to the phase shift and the visibility change in the interference pattern.
It is convenient to note that, with $P_{1,2}$ being the paths considered in \rSec{SMRMO}, 
\alg{
\!\!U(P_1)^\dagger U(P_2)=e^{\frac{\bar{E}\Delta\tau}{i\hbar}}\!\left(e^{-\frac{\Delta E\Delta\tau}{i\hbar}}\proj{g}+e^{\frac{\Delta E\Delta\tau}{i\hbar}}\proj{e}\right)\!.\!\!
\laeq{UTrTl}
}

When the attracting force of the gravity is  negligibly small, the path $\bar{P}_{1,2}$ can be approximated by a straight line.
With $w$ being the width of the interferometer, $r(\phi)=w/(2\cos{\phi})$. 
If the path is sufficiently long in both directions compared to $w$, we thus obtain, from \req{DeltaTauGL},
\alg{
\Delta\tau
&=
\frac{E}{mc^2}\frac{4GJ}{c^4}\int_{-\frac{\pi}{2}}^{\frac{\pi}{2}}\frac{2\cos{\phi}}{w}d\phi+T\cdot O\left(\frac{GM}{c^2r}\frac{GJ}{c^3r^2}\frac{v}{c}\right)\\
&=
\frac{E}{mc^2}\frac{16GJ}{c^4w}+T\cdot O\left(\frac{GM}{c^2r}\frac{GJ}{c^3r^2}\frac{v}{c}\right).
\laeq{DeltaTauGL2}
}
Here, $E$ is the total energy of the particle evaluated with respect to the Schwarzschild component of the metric, $v$ is the velocity of the particle, and $T$ is the time scale of the experiment that is in the same order as $\Delta\tau$.
The ratio $E/mc^2$ is independent of the rest mass, and is invariant in time due to the energy conservation.
It is given by
\alg{
\frac{E}{mc^2}
=
1+\frac{1}{2}\frac{v^2}{c^2}-\frac{GM}{c^2r}+o\left(\frac{v^2}{c^2},\frac{GM}{c^2r}\right).
}
Hence, in the leading order, the ratio remains equal to $K:=1+v_0^2/2c^2-GM/c^2r_0$, where $v_0$ is the velocity of the particle at a distance $r_0$ from the source mass.
The proper time difference \req{DeltaTauGL2} is thus given by
\alg{
\Delta\tau
=
\frac{16GJK}{c^4w}+T\cdot o\left(\frac{v^2}{c^2},\frac{GM}{c^2r},\frac{GJ}{c^3r^2}\frac{v}{c}\right).
\laeq{DeltaTauGL3}
}
From this expression, it becomes clear under what conditions the higher-order terms can be neglected: this is the case if the velocity of the particle is sufficiently small, $\frac{v^2}{c^2}\ll1$, and the particle remains sufficiently far from the source mass so that both the gravitational potential and the frame-dragging contributions are small, i.e., $\frac{GM}{c^2r},\frac{GJ}{c^3r^2}\frac{v}{c}\ll1$.
 We ignore the higher order terms and, for convenience, define
\alg{
w':=\frac{1}{\Delta\tau}=\frac{c^4w}{16GJK}.
\laeq{dfnwp}
}

In \rSec{DPSex}, we consider the case in which the source mass is treated as classical. Both the direction of the rotation axis and the rotation frequency are definite and fixed, with the axis oriented perpendicular to the arms of the interferometer. The resulting effect of the gravitational field on the interference pattern will be analyzed.
In \rSec{GMEex}, we turn to the case where the source mass is quantum. While the rotation frequency remains fixed, the axis of rotation is placed in a superposition of opposite directions. In this setting, we will evaluate the entanglement between the rotational degree of freedom of the source mass and the path and internal degrees of freedom of the clock particle.
In both cases, we consider the setup depicted in  \rFig{A}, except that in the latter case the rotation of the source mass is in a superposition. The analysis is based on Eqs.~\req{UTrTl}, \req{DeltaTauGL3} and \req{dfnwp}.

\subsection{Interferometric Visibility Experiment}
\lsec{DPSex}

\begin{figure}[t]
\includegraphics[bb={0 0 410 207}, scale=0.6]{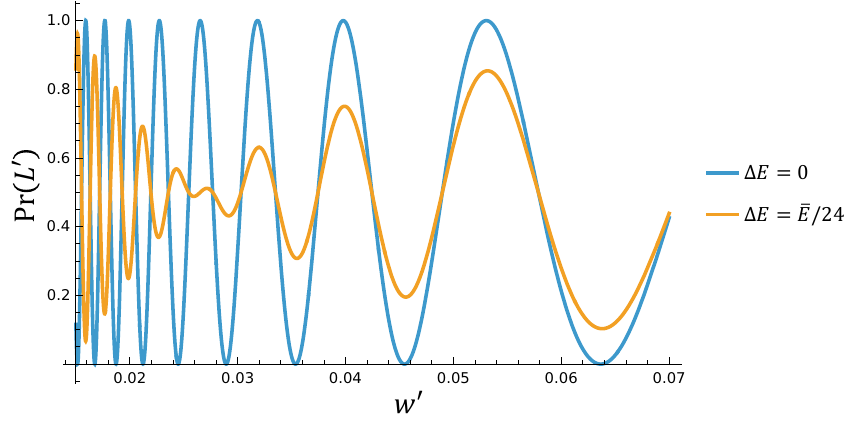}
\caption{The interference pattern predicted from Eqs.~\req{alphadfn} and \req{PrL} is shown. The vertical axis represents the probability that the detector at the left port clicks, and the horizontal axis denotes the width of the interferometer $w'$ (see Eq.~\req{dfnwp}). The blue and orange lines correspond to $\Delta E = 0$ and $\Delta E = \bar{E}/24$, respectively.}\label{fig:IV}
\end{figure}

Suppose that the internal state of the clock particle is initially in an equal superposition of the two states: $|\xi_0\rangle=\frac{|e\rangle+|g\rangle}{\sqrt{2}}$.
Right after the first beamsplitter, the whole state of the particle, namely that of the path and internal degrees of freedom, is given by
\alg{
|\Psi_{\rm ini}\rangle=\frac{1}{\sqrt{2}}(|L\rangle+|R\rangle)\otm|\xi_0\rangle.
}
As the particle propagates along the paths, it interacts with the gravitational field produced by the source mass.
Right before the second beam splitter, the state becomes
\alg{
|\Psi_{f}\rangle=
&
\frac{1}{\sqrt{2}}\left(|L\rangle\otm|\xi_1\rangle+|R\rangle\otm|\xi_2\rangle\right),
}
where
\alg{
|\xi_{1,2}\rangle:=U(P_{1,2})|\xi_0\rangle,
\laeq{xi12dfn}
}
which is transformed by the second beamsplitter into
\alg{
|\Psi_{f}'\rangle=
\frac{1}{\sqrt{2}}|L'\rangle\otm
|\eta_+\rangle
+\frac{1}{\sqrt{2}}|R'\rangle\otm
|\eta_-\rangle,
}
where $|\eta_\pm\rangle$ are unnormalized vectors defined by 
$
|\eta_\pm\rangle
=
(|\xi_1\rangle\pm|\xi_2\rangle)/\sqrt{2}
$.
Using \req{UTrTl}, the inner product of the clock states is calculated to be
\alg{
\langle\xi_1|\xi_2\rangle
=\mathcal{V}_{\Delta E,\Delta\tau}\exp(\bar{E}\Delta\tau/i\hbar)\laeq{xi12}
}
and
\alg{
\inpro{\eta_\pm}{\eta_\pm}&=1\pm\mathcal{V}_{\Delta E,\Delta\tau}\cos(\bar{E}\Delta\tau/\hbar),\laeq{etapm1}\\
\inpro{\eta_\pm}{\eta_\mp}&=\pm i\mathcal{V}_{\Delta E,\Delta\tau}\sin(\bar{E}\Delta\tau/\hbar),\laeq{etapm2}
}
where the visibility parameter $\mathcal{V}_{\Delta E,\Delta\tau}$ is given by
\alg{
\mathcal{V}_{\Delta E,\Delta\tau}:=\cos\left(\frac{\Delta E\Delta\tau}{\hbar}\right).
\laeq{alphadfn}
}
Thus, the detection probabilities are given by
\alg{
{\rm Pr}(L')
&=\frac{1}{2}\left(1+\mathcal{V}_{\Delta E,\Delta\tau}\cos\left(\frac{\bar{E}\Delta\tau}{\hbar}\right)\right),
\laeq{PrL}\\
{\rm Pr}(R')
&=\frac{1}{2}\left(1-\mathcal{V}_{\Delta E,\Delta\tau}\cos\left(\frac{\bar{E}\Delta\tau}{\hbar}\right)\right).
\laeq{PrR}
}
The above expression shows that the interference pattern indicates an amplitude modulation which is periodic in the inverse of $w$ whenever $\Delta E,\Delta\tau\neq0$, as shown in \rFig{IV}.

As discussed in \cite{zych2011quantum} for the case of a homogeneous gravitational field, the amplitude modulation can be interpreted as arising from the complementarity between path interference and the which-path information. Since proper time flows at different rates along different paths, the which-path information is encoded in the internal state of the clock particle, provided its initial state is a pure state. However, as pointed out in \cite{pikovski2017time,roura2020gravitational}, the same amplitude modulation occurs even when the internal state of the clock particle is a mixture of energy eigenstates. In fact, whether the internal state of the atom is in a superposition or a mixture of energy eigenstates does not affect the interference pattern, unless the clock state is read out or a transition between eigenstates takes place. Similar to the schemes proposed in \cite{zych2011quantum, roura2020gravitational}, and in contrast to the one in \cite{ufrecht2020atom}, our approach does not involve transitions between internal states. 
 Note, however, that the clock state could, in principle, be read out using a standard Ramsey spectroscopy procedure (see Section II B of \cite{roura2020gravitational}).


\subsection{GIE Experiment}
\lsec{GMEex}

When the source mass is in a quantum superposition of opposite rotational directions, the spacetime curvature affects the quantum clock particle in a way that leads to entanglement between the particle and the source mass.
For simplicity, we assume that the source mass has two rotational states, namely, clockwise and counterclockwise, which  will be denoted by $|\!\uparrow\rangle$ and $|\!\downarrow\rangle$, respectively.

Following the formalism presented in \cite{giacomini2020einstein}, we assume that (i) macroscopically distinguishable states of the gravitational field are assigned orthogonal quantum state vectors; (ii) each well-defined gravitational field is described by general relativity; and (iii) the quantum superposition principle holds for such gravitational fields.
Additionally, we assume that the back action of the clock particle to the gravitational field is negligible. 

 The amount of entanglement will then be calculated using the quantities introduced in the previous subsection.

\subsubsection{ Description of The Protocol}

For the preparation of the superposition state of rotational directions, we exploit the protocol proposed in \cite{higgins2024truly}.
Here, the source mass is assumed to be a particle that has an electric dipole moment represented by $\{|0\rangle,|1\rangle\}$ and a large magnetic dipole moment.
At the first step, the electric dipole moment is prepared in a superposition state $(|0\rangle+|1\rangle)/\sqrt{2}$.
Then an electric field is applied, by which the orientation $\theta$ of the particle becomes entangled so that the state is $(|0\rangle|\theta=-\theta_0\rangle+|1\rangle|\theta=\theta_0\rangle)/\sqrt{2}$. In particular, we choose $\theta_0=\pi/2$.
Next, an alternating magnetic field is applied so that the particle starts rotating around a given axis with frequency $\omega_0$.
The state will become $(|0\rangle|\theta=-\pi/2,\omega=\omega_0\rangle+|1\rangle|\theta=\pi/2,\omega=\omega_0\rangle)/\sqrt{2}\equiv(|0\rangle|\downarrow\rangle+|1\rangle|\uparrow\rangle)/\sqrt{2}$.
Taking the state of the gravitational field into account, this process realizes the transformation
\alg{
&\frac{|0\rangle+|1\rangle}{\sqrt{2}}|\theta=0,\omega=0\rangle|g_0\rangle
\rightarrow
\nn\\
&
\quad
\frac{1}{\sqrt{2}}\left(|0\rangle|\!\uparrow\rangle|g_\uparrow\rangle+|1\rangle|\!\downarrow\rangle|g_\downarrow\rangle\right)
\equiv
\frac{|\!\Uparrow\rangle+|\!\Downarrow\rangle}{\sqrt{2}},
\laeq{gracial}
}
where $g_0$, $g_\uparrow$ and $g_\downarrow$ are the states of the gravitational field generated by the source mass that is at rest, rotating clockwise or counterclockwise, respectively. Exactly the same procedure in the reverse order will take the state back to the initial state.

  Once the source mass has been prepared in the superposition state, the clock particle enters the interferometer. Immediately after the first beam splitter, the total state of the system, consisting of the source, the gravitational field, the path degree of freedom, and the internal clock, is in the state
\alg{
|\Psi_{\rm ini}\rangle=\left(\frac{|\!\Uparrow\rangle+|\!\Downarrow\rangle}{\sqrt{2}}\right)_ {\!\! SG}\otm\left(\frac{|L\rangle+|R\rangle}{\sqrt{2}}\right)_{\!\! P}\otm|\xi_0\rangle_C,
}
where subscripts $S$, $G$, $P$ and $C$ denote the source, the gravitational field, the path and clock degrees of freedom, respectively.
Right before the second beam splitter, the state becomes
\alg{
|\Psi_{\rm fin}\rangle=
&
\frac{1}{2}|\!\Uparrow\rangle\otm\left(|L\rangle\otm|\xi_1\rangle+|R\rangle\otm|\xi_2\rangle\right)
\nn\\
&\!\!+\frac{1}{2}|\!\Downarrow\rangle\otm\left(|L\rangle\otm|\xi_2\rangle+|R\rangle\otm|\xi_1\rangle\right)\!,\!
\laeq{psifin}
}
where $\xi_{1,2}$ are given by \req{xi12dfn}.
This state will be further transformed by the second beamsplitter.
By the reverse process of \req{gracial}, the field degree of freedom will be disentangled from the particles.
After these procedures, the state of the electric dipole moment of the source particle, and the path and the internal degrees of the clock particle, is 
\alg{
\!|\Psi_{\rm fin}'\rangle=
\frac{1}{\sqrt{2}}\left(|{\bf +}\rangle\otm|L'\rangle\otm|\eta_+\rangle+|-\rangle\otm|R'\rangle\otm|\eta_-\rangle\right)\!,\!
\laeq{sfsbm}
}
where $|\pm\rangle:=(|0\rangle\pm|1\rangle)/\sqrt{2}$.

\subsubsection{Evaluation of Entanglement}

Let us evaluate the amount of entanglement of the state $\Psi'_{\rm fin}$.
Note that $\Psi'_{\rm fin}$ is a pure state on $S$, $P$ and $C$.
Thus, the entanglement between the source mass and the clock particle is quantified by the entanglement entropy \cite{bennett1996concentrating}.
 From \req{etapm1} and \req{etapm2}, it is given by
\alg{
\mathcal{E}^{S|PC}(\Psi'_{\rm fin})=h\left(\frac{1+\mathcal{V}_{\Delta E,\Delta\tau}\cos\left(\frac{\bar{E}\Delta\tau}{\hbar}\right)}{2}\right).
\laeq{ESPC}
}
Here, $h$ is the binary entropy defined by  $h(x)=-x\log_2{x}-(1-x)\log_2{(1-x)}$ for $0\leq x\leq1$, which is concave and symmetric around $x=1/2$, with a minimum value $h(0)=h(1)=0$ and maximum $h(1/2)=1$.
This shows that (i) the maximum amount of entanglement does not depend on $\Delta E$ because $\mathcal{E}_E^{S|PC}(\Psi'_{\rm fin})=1$ whenever $\cos(\frac{\bar{E}\Delta\tau}{\hbar})=0$, (ii) the minimum amount of entanglement is larger for $\Delta E\neq0$ than $\Delta E=0$, and that (iii) there is a modulation of the amount of entanglement that is periodic in the inverse of $w$,  whenever $\Delta E\neq0$, due to $\mathcal{V}_{\Delta E,\Delta\tau}$. 

\begin{figure}[t]
\includegraphics[bb={0 0 422 205}, scale=0.6]{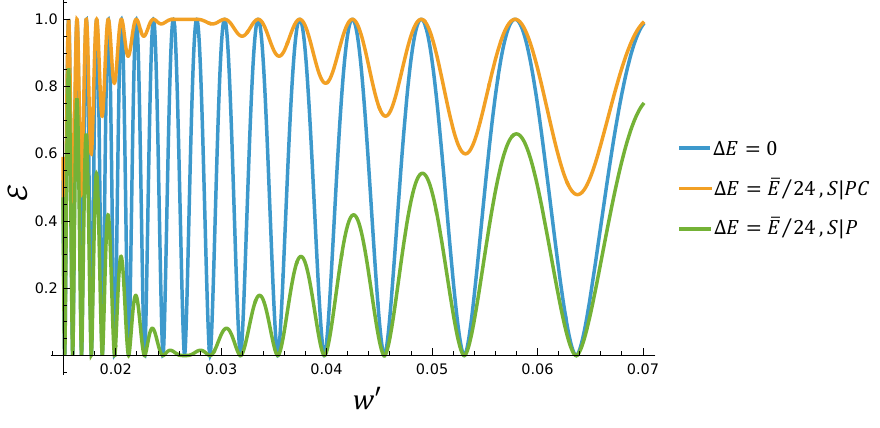}
\caption{The amount of entanglement predicted from Eqs.~\req{ESPC} and \req{ESP} is shown. The horizontal axis denotes the width of the interferometer $w$. The blue line represents the entanglement for the case of $\Delta E = 0$, while the orange and green lines correspond to $\mathcal{E}^{S|PC}$ and $\mathcal{E}^{S|P}$, respectively, for $\Delta E = \bar{E}/24$.
}\label{fig:GME}
\end{figure}

 We also evaluate the entanglement between the source mass and the path degree of freedom of the clock particle, ignoring the clock degree of freedom. 
The reduced state of $\Psi'_{\rm fin}$ on system $SP$ is in general a mixed state because $\Psi'_{\rm fin}$ is entangled between $SP$ and $C$.
In such a case, it is necessary to adopt measures of entanglement that applies to general mixed states, such as the entanglement of formation \cite{wootters1998entanglement}.
 Noting \req{etapm1} and \req{etapm2},
the entanglement of formation between the source mass and the path degree of freedom is evaluated as
\alg{
\mathcal{E}^{S|P}(\Psi'_{\rm fin})=h\left(\frac{1+\sqrt{1-\mathcal{V}_{\Delta E,\Delta\tau}^2\sin^2\left(\frac{\bar{E}\Delta\tau}{\hbar}\right)}}{2}\right)\!,\!
\laeq{ESP}
}
see \rFig{GME}.
It is straightforward that
\alg{
\mathcal{E}^{S|PC}(\Psi'_{\rm fin})=\mathcal{E}^{S|P}(\Psi'_{\rm fin})
=h\left(\frac{1+\cos\left(\frac{\bar{E}\Delta\tau}{\hbar}\right)}{2}\right)
}
whenever $\Delta E=0$.

 These results indicate that, from the standpoint of maximizing bipartite entanglement, employing a quantum clock ($\Delta E\neq0$) does not provide an advantage.
In particular, the amount of entanglement between the source mass ($S$) and the path degree of freedom of the particle ($P$) is smaller in the case of $\Delta E\neq0$ than in the case of $\Delta E=0$.
This is because, in the former case, the entanglement between $C/SP$ makes the reduced state on $SP$ a mixed state, which leads to decrease of entanglement between $S/P$.
Nevertheless, in the context of quantum equivalence principle, there is still an advantage in employing a particle whose internal degree of freedom evolves in time (see \rSec{IEQPGME}).

\subsubsection{ Measurement Requirement}

We briefly analyze a measurement requirement for witnessing the generated entanglement.
In order to witness {\it tripartite} entanglement among $S$, $P$ and $C$, measurements should be performed not only on the source mass and the path degree of freedom of the clock particle, but also on the  clock degree of freedom of it.
This requires a direct readout of the  clock, which is in general experimentally more demanding than 
observing the path interference (though it is possible by Ramsey spectroscopy, see Section II B of \cite{roura2020gravitational}).
Therefore, we only consider a measurement requirement for witnessing entanglement between $S/P$.

The entanglement in the state $\Psi'_{\rm fin}$ between $S$ and $P$ can be witnessed \cite{bose2017spin} by measuring the correlation functions
\alg{
\mathcal{W}_\pm
:=
\left|\langle X^S\otm X^P\pm Y^S\otm Z^P\rangle\right|
,
}
where
\alg{
X^S\equiv |+\rangle\!\langle+|-|-\rangle\!\langle-|,\quad
Y^S\equiv -i|0\rangle\!\langle1|+i|1\rangle\!\langle0|
}
and
\alg{
X^P\equiv |L'\rangle\!\langle L'|-|R'\rangle\!\langle R'|,\quad
Z^P\equiv |L\rangle\!\langle L|-|R\rangle\!\langle R|.
}
Indeed, one can show that $\mathcal{W}_\pm>1$ only if the state is entangled.
Using Eqs.~\req{xi12}-\req{etapm2}, \req{psifin} and \req{sfsbm}, the correlation function is calculated to be
\alg{
\mathcal{W}_\pm=1\pm\mathcal{V}_{\Delta E,\Delta\tau}\sin\left(\frac{\bar{E}\Delta\tau}{\hbar}\right).
}
Thus, the entanglement will be witnessed whenever $|\mathcal{V}_{\Delta E,\Delta\tau}\sin(\bar{E}\Delta\tau/\hbar)|\neq0$, as expected from \req{ESP}.
Protocols that are more sensitive to the generated entanglement could be devised based on, e.g.,  \cite{chevalier2020witnessing,guff2022optimal}.

\section{Analysis based on Quantum Equivalence Principle}
\lsec{qEP}

In this section, we analyze implications of the quantum equivalence principle (QEP) in the experiments presented in the previous sections.
In \rSec{FQEP}, we  extend the QEP of \cite{zych2018quantum} to nonstatic stationary spacetimes.
In \rSec{TQEP}, we show that the  extended QEP can be tested in the interferometric visibility experiment.
 
In \rSec{IEQPGME}, we propose an adaptation of the generalized QEP to the test of GIE.
In \rSec{impQEPGME}, we discuss conclusions that can be drawn from the possible violations of the generalized QEP in the above experiments.

\subsection{Formulation of Quantum Equivalence Principle in post-Newtonian Classical Spacetime}
\lsec{FQEP}

As we have shown in \rSec{qCSS} (see also Eq.~\req{taudt} in \rApp{pathint}), the Routhian of the particle in stationary spacetime is approximately given by
\alg{
\hat{R}=\hat{H}_{rest}\left(
1-\frac{v^2}{2c^2}+\frac{\Phi}{c^2}-\sum_{i=1}^3\frac{g_{0i}}{cg_{00}}\frac{dx^i}{dt}\right)\!.
\laeq{routhEP1}
}
In the case of the static spacetime in the Newtonian limit, where the last term in the above  are absent, a phenomenological model for the violation of quantum equivalence principle is provided in \cite{zych2018quantum}.
In terms of the Routhian representation, it is represented as
\alg{
\tilde{R}_{st}=\hat{H}_{r}
-\hat{H}_{i}\frac{v^2}{2c^2}+\hat{H}_{g}\frac{\Phi}{c^2},
\laeq{routhEP2}
}
where $\hat{H}_{r}$, $\hat{H}_{i}$ and $\hat{H}_{g}$ are the internal Hamiltonians that correspond to the rest mass, the inertial mass and the (passive) gravitational mass, respectively. 
The quantum equivalence principle is then formulated as the equality $\hat{H}_{r}=\hat{H}_{i}=\hat{H}_{g}$.
The equality asserts that not only the spectra but also the eigenvectors of the Hamiltonians are all equal, which distinguishes the quantum equivalence principle from the classical one.

We  extend this model to the nonstatic stationary spacetime.
We consider a test theory in which the Routhian is represented as
\alg{
\tilde{R}=\hat{H}_{r}
-\hat{H}_{i}\frac{v^2}{2c^2}+\hat{H}_{g}\frac{\Phi}{c^2}-\hat{H}_{f}\sum_{i=1}^3\frac{g_{0i}}{cg_{00}}\frac{dx^i}{dt},
}
where $\hat{H}_{f}$ is the internal Hamiltonian that describes the way how the particle is affected by the frame dragging effect.
Contrary to $\hat{H}_{i}$ and $\hat{H}_{g}$ that have clear interpretations in Newtonian mechanics, the Hamiltonian $\hat{H}_{f}$, which has a genuinely post-Newtonian origin, does not have a clear interpretation in the context of the original formulation of the Einstein's equivalence principle.
Nevertheless, it is possible to formulate the quantum equivalence principle in this model as $\hat{H}_{r}=\hat{H}_{i}=\hat{H}_{g}=\hat{H}_{f}$.
Indeed, if the Lagrangian of a classical particle without internal degree of freedom is given by
\alg{
\tilde{L}=-m_{r}c^2
+\frac{m_{i}v^2}{2}-m_{g}\Phi+m_{f}c\sum_{i=1}^3\frac{g_{0i}}{g_{00}}\frac{dx^i}{dt},
}
and if we demand that the trajectory of the particle is determined only in terms of the metric even for nonstatic spacetimes, it must hold that $m_{r}=m_{i}=m_{g}=m_{f}$.
For the simplicity of analysis, we assume that the  QEP is valid at the Newtonian limit, i.e., $\hat{H}_{r}=\hat{H}_{i}=\hat{H}_{g}=:\hat{H}_N$.
The  extended QEP is then represented as $\hat{H}_{N}=\hat{H}_{f}$.
   In particular, it implies commutativity of the two Hamilonians $[\hat{H}_{N},\hat{H}_{f}]=0$.
The Routhian for the test theory is
\alg{
\!\!\tilde{R}=\hat{H}_{N}\left(1-\frac{v^2}{2c^2}+\frac{\Phi}{c^2}\right)-\hat{H}_{f}\sum_{i=1}^3\frac{g_{0i}}{cg_{00}}\frac{dx^i}{dt}.\!\!
\laeq{testR}
}
The phase accumulation corresponding to \req{routhian} is then given by
\alg{
\int_{\rm ini}^{\rm fin}\tilde{R}dt=&\hat{H}_{N}\int_{\rm ini}^{\rm fin}\left(1-\frac{v^2}{2c^2}+\frac{\Phi}{c^2}\right)dt\nn\\
&\quad-\hat{H}_{f}\int_{\rm ini}^{\rm fin}\sum_{i=1}^3\frac{g_{0i}}{cg_{00}}dx^i.
\laeq{testRT}
}

\subsection{Test of Quantum Equivalence Principle in Interferometric Visibility Experiment}
\lsec{TQEP}

An experimental test of the  QEP in interferometric visibility setup has been proposed in \cite{zych2017quantum,rosi2017quantum,zych2018quantum} for the case of homogeneous gravitational field.
Based on the test theory represented by the  Newtonian Routhian \req{routhEP2}, it was shown there that the violation of QEP could be detected as a change of the visibility in the interference pattern.
We here extend this model to the case of inhomogeneous nonstatic stationary spacetime.

\begin{figure}[t]
\includegraphics[bb={0 0 391 199}, scale=0.6]{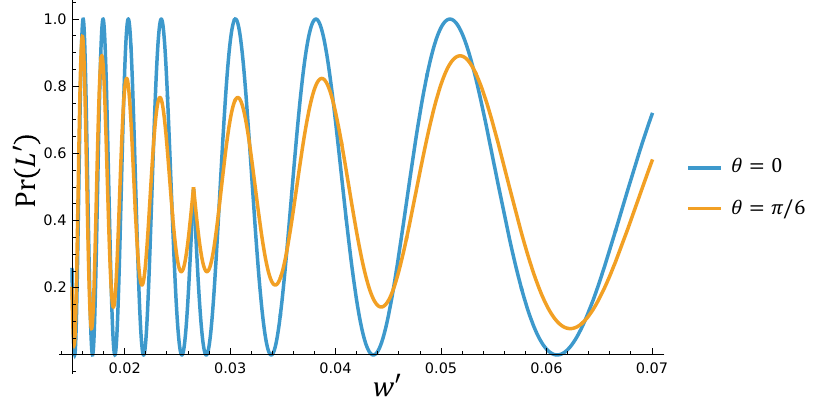}
\caption{The interference pattern predicted from Eqs.~\req{visQEP} and \req{PrLQEP} is shown. The vertical axis represents the probability that the detector at the left port clicks, and the horizontal axis denotes the width of the interferometer $w'$.  The energy eigenvalues are chosen to be $E_g'=E_g$, $E_e'=E_e$ and $\Delta E'=\Delta E=\bar{E}/24$. The blue and orange lines correspond to $\theta = 0$ and $\theta = \pi/6$, respectively.}\label{fig:IVep}
\end{figure}

We consider again the quantum clock interferometry setup described in \rSec{DPSex}. 
Instead of \req{routhEP1}, which is based on the model presented in \rSec{qCSS}, we now adopt the test theory \req{testR}, which results in the time integral of the Routhian given in \req{testRT}. 
 Furthermore, in contrast to the previous case, we now take the initial state of the clock degree of freedom, $|\chi_0\rangle$, to be one of the eigenstates of the Hamiltonian $\hat{H}_{N}$.
In the same way as \req{sfsbm}, the state after the second beamsplitter is
\alg{
|\Psi_{\rm fin}'\rangle=
\frac{1}{\sqrt{2}}|L'\rangle\otm
|\zeta_+\rangle
+\frac{1}{\sqrt{2}}|R'\rangle\otm
|\zeta_-\rangle,
}
where $|\zeta_\pm\rangle$ are unnormalized vectors defined by 
$
|\zeta_\pm\rangle
=
(|\chi_1\rangle\pm|\chi_2\rangle)/\sqrt{2}
$,
and $|\chi_{1,2}\rangle=U(P_{1,2})|\chi_0\rangle$.

 Let the eigen decomposition of the Hamiltonian $\hat{H}_{f}$ be $\hat{H}_{f}=E_{g}'\proj{g'}+E_e'\proj{e'}$. The initial state $|\chi_0\rangle$ is expanded in terms of the eigenstates $\{|g'\rangle,|e'\rangle\}$ as $|\chi_0\rangle=\cos\theta|g'\rangle+e^{i\varphi}\sin\theta|e'\rangle$. 
The commutativity of the Hamiltonians $[\hat{H}_{N},\hat{H}_{f}]=0$ is then equivalent to $\theta=\ell\pi/2\:(\ell\in\mbb{N})$.

We now evaluate the probability of detecting the clock particle at each detector to examine how the possible violation of the extended QEP can be observed in the interference pattern.
We assume that $\|[\hat{H}_{N},\hat{H}_{f}]\|\ll\|\hat{H}_{N}\|$, in which case the relative evolution operator becomes
\alg{
U(P_1)^\dagger U(P_2)
&\approx
\exp{\left(\frac{\hat{H}_{f}}{i\hbar}\int_{\bar{P}}\sum_{i=1}^3\frac{2g_{0i}}{cg_{00}}dx^i\right)}
\\
&=
\exp{\left(\frac{\hat{H}_{f}\Delta\tau}{i\hbar}\right)}\\
&=e^{\frac{E_{g}'\Delta\tau}{i\hbar}}\proj{g'}+e^{\frac{E_e'\Delta\tau}{i\hbar}}\proj{e'},
}
where $\Delta\tau$ is given by \req{DeltaTauGL3}.
As we prove in \rApp{prfvisQEP}, we have
\alg{
\!\inpro{\chi_1}{\chi_2}=
\cos^2\theta\exp\left(\!\frac{E_{g}'\Delta\tau}{i\hbar}\!\right)\!+\sin^2\theta\exp\left(\!\frac{E_e'\Delta\tau}{i\hbar}\!\right)\!
\laeq{visQEP1}
}
and
\alg{
\inpro{\zeta_\pm}{\zeta_\pm}&=1\pm\mathcal{V}_{\Delta E',\Delta\tau,\theta}\cos\left(\frac{(\bar{E}'+\xi)\Delta\tau}{\hbar}\right),\laeq{visQEP2}\\
\inpro{\zeta_\pm}{\zeta_\mp}&=\pm i\mathcal{V}_{\Delta E',\Delta\tau,\theta}\sin\left(\frac{(\bar{E}'+\xi)\Delta\tau}{\hbar}\right).\laeq{visQEP3}
}
Here, the visibility parameter is expressed as
\alg{
\mathcal{V}_{\Delta E',\Delta\tau,\theta}
=
\sqrt{1-\sin^22\theta\sin^2\left(\frac{\Delta E'\Delta\tau}{\hbar}\right)}
\laeq{visQEP}
}
and $\xi$ is given by
\alg{
\tan{\left(\frac{\xi\Delta\tau}{\hbar}\right)}=-\cos2\theta\tan\left(\frac{\Delta E'\Delta\tau}{\hbar}\right).
\laeq{visQEP5}
}
The detection probabilities are thus given by
\alg{
{\rm Pr}(L')
&=\frac{1}{2}\left(1+\mathcal{V}_{\Delta E',\Delta\tau,\theta}\cos\left(\frac{(\bar{E}'+\xi)\Delta\tau}{\hbar}\right)\right),
\laeq{PrLQEP}
\\
{\rm Pr}(R')
&=\frac{1}{2}\left(1-\mathcal{V}_{\Delta E',\Delta\tau,\theta}\cos\left(\frac{(\bar{E}'+\xi)\Delta\tau}{\hbar}\right)\right),
}
and is plotted in \rFig{IVep}.

The presence of the amplitude modulation in the interference pattern is an indication of the violation of the quantum equivalence principle.
Indeed, it can be seen from \req{visQEP} and \req{DeltaTauGL3} that there is an amplitude modulation that is periodic in $1/w$, whenever $\theta\neq\ell\pi/2\:(\ell\in\mbb{N})$, that is, $[\hat{H}_N,\hat{H}_f]\neq0$.
Thereby the violation of the quantum equivalence principle in nonstatic spacetimes could, in principle, be detected in a manner similar to that discussed in Ref.~\cite{zych2018quantum}.

\subsection{ Adaptation of Quantum Equivalence Principle to GIE Experiment}
\lsec{IEQPGME}

By generalizing the model defined above, one can formulate the question of  whether the gravity-induced entanglement observed in the experiment presented in \rSec{DPSex} satisfies the extended QEP. Using the test theory  represented by Eq.~\req{testR}, it is possible to theoretically model a violation of the  extended QEP in the context of gravity-induced entanglement. In this model, it is shown that if the  extended QEP does not hold, the magnitude of the generated entanglement exhibits an amplitude modulation. Therefore, by detecting the presence or absence of this modulation, one can experimentally test whether the generated gravity-induced entanglement  is consistent with the equivalence principle.

As in \rSec{GMEex}, we make the following assumptions: (i) orthogonal quantum state vectors are assigned to macroscopically distinguishable states of the gravitational field; (ii) each gravitational field configuration obeys the laws of general relativity; (iii) the superposition principle applies to such gravitational fields; and (iv) the back action of the clock particle on the gravitational field is negligible.
The extended QEP then requires that, for each configuration of the gravitational field, its interaction with the clock particle satisfies the  extended QEP described in \rSec{FQEP}.
This formulation remains consistent even when the gravitational field is in a quantum superposition, in line with the ideas presented in Ref.~\cite{giacomini2020einstein}.

\begin{figure}[t]
\includegraphics[bb={0 0 411 200}, scale=0.6]{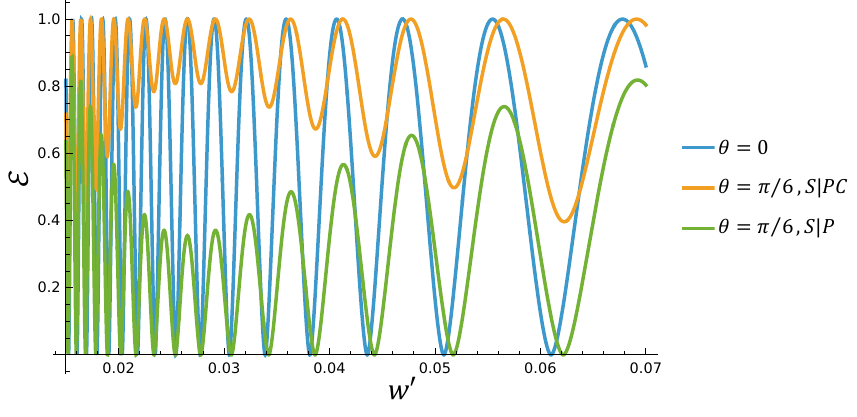}
\caption{The amount of entanglement predicted from Eqs.~\req{ESPCQEP} and \req{ESPQEP} is shown. The horizontal axis denotes the width of the interferometer $w'$. The blue line represents the entanglement for the case of $\theta = 0$, while the orange and green lines correspond to  $\mathcal{E}^{S|PC}$ and $\mathcal{E}^{S|P}$, respectively, for the case of $\theta = \pi/6$.
}\label{fig:GMEep}
\end{figure}

We now consider the case where the source mass is in a superposition of the opposite rotation directions.
After the second beamsplitter, the state is
\alg{
\!\!|\Psi_{\rm fin}'\rangle=
\frac{1}{\sqrt{2}}\left(|+\rangle\otm|L'\rangle\otm|\zeta_+\rangle+|-\rangle\otm|R'\rangle\otm|\zeta_-\rangle\right)\!.\!
}
The entanglement between the source mass and the clock particle,  incorporating and ignoring the clock degree of freedom, are evaluated as
\alg{
\mathcal{E}_E^{S|PC}(\Psi'_{\rm fin})=h\left(\frac{1+\mathcal{V}_{\Delta E',\Delta\tau,\theta}\cos\left(\frac{(\bar{E}'+\xi)\Delta\tau}{\hbar}\right)}{2}\right)
\laeq{ESPCQEP}
}
and
\alg{
\!\!\mathcal{E}_F^{S|P}(\Psi'_{\rm fin})=h\left(\frac{1+\sqrt{1-\mathcal{V}_{\Delta E',\Delta\tau,\theta}^2\sin^2\left(\frac{(\bar{E}'+\xi)\Delta\tau}{\hbar}\right)}}{2}\right)\!\!,\!
\laeq{ESPQEP}
}
respectively, and are depicted in \rFig{GMEep}.
The result shows that the oscillation of the generated entanglement indicates the amplitude modulation whenever $\mathcal{V}_{\Delta E',\Delta\tau,\theta}\neq1$, that is, when the  extended QEP is violated.

\subsection{Implication of Quantum Equivalence Principle to GIE Experiment}
\lsec{impQEPGME}

We now discuss what can be inferred from possible violations of the  extended QEP in the interferometric visibility (IV) experiment (\rSec{TQEP}) and in the GIE experiment (\rSec{IEQPGME}). 
In particular, we examine the implications of the three possible outcomes of QEP violation across the two experiments, summarized in Table~\ref{table:QEPv}.
If no QEP violation is observed in either the IV or the GIE experiment (Case 1), the results would support the validity of the QEP, extended to include frame-dragging effects, in both classical and quantum background spacetimes.
If QEP violation is observed in both experiments (Case 2), the conclusion would be that the QEP, at least in the form considered here, does not hold in the post-Newtonian regime.
The most significant case is when no QEP violation is observed in the IV experiment but a violation {\it is} observed in the GIE experiment (Case 3). In this situation, one can conclude something about the mechanism by which gravity mediates entanglement: the observed GIE cannot be accounted for by models in which the gravitational field is described as a quantum superposition of classical spacetime geometries. Accordingly, our proposed scheme offers an experimental method capable of potentially ruling out one of the quantum-gravity models proposed to explain how GIE is generated.

\begin{table}[t]
\begin{tabular}{|l|l|l|l|}\hline
&IV&GIE&\\\hline
Case 1 & No & No & consistent with QEP\\
Case 2 & Yes & Yes & breakdown of post-Newtonian QEP\\
Case 3 & No & Yes & rules out superposed geometry models\\ \hline
\end{tabular}
\caption{ Implications of possible QEP violations in interferometric visibility and gravity-induced entanglement are summarized.}
\label{table:QEPv}
\end{table}

\section{Discussion}
\lsec{discussion}

In this work, we have proposed and theoretically analyzed two experimental schemes aimed at probing post-Newtonian gravitational effects using quantum clock interferometry. Our focus was on the frame-dragging effect generated by a rotating source mass, particularly its effect on the proper-time difference measured by a quantum clock, as well as its potential to generate gravity-induced entanglement. The experimental configurations were designed such that Newtonian contributions cancel due to symmetry, thereby isolating post-Newtonian effects as the dominant source.
Note that quantum clock interferometry is conceptually different from the atom interferometry, which has been used as a quantum probe to detect spacetime curvature \cite{roura2022quantum}.

To estimate the scale of the effect, we evaluate the visibility parameter \req{alphadfn} using concrete values. A typical quantum clock has a transition frequency of $\Delta E/\hbar \sim 10^{15}\,{\rm rad/s}$. For a setup with $w \sim 1\,{\rm mm}$ and dimensionless angular momentum $\ell \equiv J/\hbar$, the resulting phase shift is of order $\ell \cdot 10^{-60}$. Achieving a detectable shift would therefore require $\ell \sim 10^{60}$, corresponding to a planetary-scale rotating mass, which is well beyond any realistic laboratory setting!

This analysis shows that while quantum clock interferometry offers a conceptually compelling method to isolate post-Newtonian gravitational signatures, the magnitude of these effects is exceedingly small within any parameter regime accessible to tabletop experiments. This suppression arises from the fact that the frame-dragging effect scales with the angular momentum of the rotating source and inversely with the fourth power of the speed of light. Even for relatively large laboratory-scale sources, the resulting proper-time differences or entanglement remain negligibly small. Therefore, the method proposed in this paper should be regarded primarily as a Gedankenexperiment. Nevertheless, our scheme provides a useful starting point for exploring the detectability of post-Newtonian quantum gravitational effects in experiments that may become feasible with near-future quantum technologies.

Future work will focus on establishing a field-theoretic description of gravity-induced entanglement based on linearized quantum gravity in our setup, as done in the path protocol \cite{christodoulou2023locally,martin2023gravity} and the oscillator protocol \cite{bose2022mechanism,bengyat2024gravity}. Another direction is to further investigate the role of the quantum equivalence principle in the context of gravity-induced entanglement.

\begin{acknowledgments}
The author thanks Ofek Bengyat, Andrea Di Biagio, Caslav Brukner, Carlo Cepollaro, Luciano Petruzziello and Jiro Soda for useful discussions.
This work is supported by MEXT Quantum Leap Flagship Program (MEXT QLEAP), Grant No.~JPMXS0120319794.
\end{acknowledgments}

\bibliography{/Users/eyuriwakakuwa/Dropbox/DropTop/latexfiles/bibliography.bib}

\appendix

\begin{widetext}

\section{Derivation of the path integral formula}
\lapp{pathint}

In this section, we provide a detailed proof that the time evolution of a quantum clock particle is described by a path-integral formula as
\alg{
\inpro{\xi_{\rm fin},q_{\rm fin};t_{\rm fin}}{\xi_{\rm ini},q_{\rm ini};t_{\rm ini}}
=
\bra{\xi_{\rm fin}}\int\mathscr{D}q\sqrt{\gamma(q)}\exp{\left(\frac{\hat{H}_{rest}\tau}{i\hbar}\right)}\ket{\xi_{\rm ini}},
\laeq{propagator}
}
where $(q_{\rm ini};t_{\rm ini})$ and $(q_{\rm fin};t_{\rm fin})$ denote spacetime points, and $\xi_{\rm ini}$ and $\xi_{\rm fin}$ denote internal states of the particle.
The integral $\int\mathscr{D}q$ is taken over all timelike trajectories from $(q_{\rm ini},t_{\rm ini})$ to $(q_{\rm fin},t_{\rm fin})$, with $\gamma$ being the determinant of the spatial components of the metric. $\hat{H}_{rest}$ is the rest Hamiltonian of the particle, and $\tau=\int_{\rm ini}^{\rm fin}d\tau$ is the proper time along each trajectory.
To prove the above, we impose the following assumptions:
\begin{enumerate}
\renewcommand{\theenumi}{\roman{enumi}}
\renewcommand{\labelenumi}{(\theenumi)}
\item {\it Weak gravity}: Gravity is sufficiently weak so that the deviation of the spacetime metric from the Minkowski metric can be treated as a small perturbation.
\item {\it Stationary spacetime}: The spacetime metric is invariant in the coordinate time.
\item {\it Slow motion}: The velocity of the particle is sufficiently small, compared to the light speed, so that the higher-order terms of $v/c$ is negligible.
\item {\it Low energy}: The energy gap of the internal (rest) Hamiltonian is sufficiently small compared to the rest mass (i.e.~ the energy of the ground state).
\end{enumerate}
We follow the standard procedure to obtain a path-integral formula from the canonical quantization (see e.g.~Chapter 9.1 of \cite{peskin2018introduction} or Section 6 of \cite{srednicki2007quantum}), except that the rest Hamiltonian is kept unintegrated when transforming the Hamiltonian representation of the action integral to the Lagrangian representation.

\subsection{Calculation of the Spacetime Metric}
\lapp{gij}

We first establish relations among metric components for stationary (i.e. invariant in the coordinate time) but nonstatic (i.e. it has non-zero $g_{0i}=g_{i0}$ components) spacetimes, which will be used to express the Hamiltonian in canonical form.
Without loss of generality, we assume that the spatial part of the metric is diagonal, i.e., $g_{ij}=0$ whenever $i\neq j$.
Indeed, the metric can be transformed in such a way by a proper transformation on the spatial part of the coordinate.
We show below that the components satisfy
\alg{
\frac{g^{0i}}{g^{00}}=\frac{g_{0i}}{g_{ii}},
\quad
g^{00}=\frac{g_{11}g_{22}g_{33}}{g}
\laeq{gii1}
}
and
\alg{
\frac{1}{g^{00}}= g_{00}-\sum_{i=1}^3\frac{g_{0i}^2}{g_{ii}},\;\;
g^{ij}=\frac{\delta_{ij}}{g_{ii}}+\frac{g_{0i}g_{0j}\prod_{k=1}^3g_{kk}}{gg_{ii}g_{jj}},\!
\laeq{gii2}
}
where $g={\rm det}(g_{\mu\nu})$.

\bprf
Since we assume that $g_{ij}=0$ for $i\neq j$, the metric tensor is of the form
\alg{
(g_{\mu\nu})=
\begin{bmatrix}
g_{00}&g_{01}&g_{02}&g_{03}\\
g_{01}&g_{11}&0&0\\
g_{02}&0&g_{22}&0\\
g_{03}&0&0&g_{33}
\end{bmatrix}.
\laeq{metric}
}
The determinant of the above matrix is given
\alg{
g:={\rm det}(g_{\mu\nu})=g_{00}g_{11}g_{22}g_{33}-g_{01}^2g_{22}g_{33}-g_{02}^2g_{33}g_{11}-g_{03}^2g_{11}g_{22}.
\laeq{metricdet}
}
The components of $g^{\mu\nu}$, which is the inverse matrix of \req{metric}, is calculated in terms of the adjugate matrix.
In particular, we have
\alg{
g^{00}=\frac{g_{11}g_{22}g_{33}}{g},
\quad
g^{01}=\frac{g_{01}g_{22}g_{33}}{g},
\quad
g^{12}=\frac{g_{01}g_{02}g_{33}}{g}
\laeq{misu}
}
and
\alg{
g^{11}=\frac{g_{00}g_{22}g_{33}-g_{02}^2g_{33}-g_{03}^2g_{22}}{g}.
}
It immediately follows from \req{misu} that
\alg{
\frac{g^{01}}{g^{00}}=\frac{g_{01}}{g_{11}}.
}
Furthermore, using \req{metricdet}, we obtain
\alg{
\frac{1}{g^{00}}=g_{00}-\frac{g_{01}^2}{g_{11}}-\frac{g_{02}^2}{g_{22}}-\frac{g_{03}^2}{g_{33}}
}
and
\alg{
gg^{11}g_{11}=g+g_{01}^2g_{22}g_{33},
}
the latter of which leads to
\alg{
g^{11}=\frac{1}{g_{11}}\left(1+\frac{g_{01}^2g_{22}g_{33}}{g}\right).
}
By a cyclic change of the indices, we obtain \req{gii1} and \req{gii2}.
\QED
\eprf

\subsection{Derivation of An Approximate Classical Hamiltonian}

We next derive the classical Hamiltonian consistent with the post-Newtonian expansion under the above metric.
Consider a classical particle with the rest mass $m$. For the moment, we assume that it has no internal degree of freedom.
We start with deriving an approximate classical Hamiltonian of the particle, based on the relativistic dissipation relation
\alg{
g^{\mu\nu}p_\mu p_\nu=-m^2c^2,
}
where $p_\mu$ is the four-momentum of the particle.
This equation can be solved as a quadratic equation with respect to $p_0(=E/c)$, with $E$ being the total energy of the particle. Noting that $g^{00}<0$, the only nonnegative solution is given by
\alg{
\frac{E}{c}
=
-\frac{g^{0i}p_i}{g^{00}}+\sqrt{\left(\frac{g^{0i}p_i}{g^{00}}\right)^2-\frac{p_ip^i+m^2c^2}{g^{00}}}.
\laeq{Htotc1}
}
Using \req{gii1}, the first term is given by
\alg{
-\frac{g^{0i}p_i}{g^{00}}
=
-\sum_{i=1}^3\frac{g_{0i}p_i}{g_{ii}}.
\laeq{gii3}
}
Let $-2\Phi/c^2:=1+g_{00}$.
Based on the conditions of weak gravity and slow motion (Conditions (i) and (iii) above), we take the approximation that includes terms up to first order in $p_jp^j/m^2c^2$, $g_{0i}$, and in $\Phi/c^2$, and neglect the higher-order terms. 
From \req{gii2} and \req{gii3}, we have
\alg{
&
\frac{1}{g^{00}}= g_{00}+o(g_{0i})=-1-\frac{2\Phi}{c^2}+o(g_{0i}), 
\\
&
\frac{p_ip^i}{m^2c^2}=\frac{p_ip_jg^{ij}}{m^2c^2}=\frac{1}{m^2c^2}\sum_{i=1}^3\frac{p_i^2}{g_{ii}}+o(g_{0i})
}
and
\alg{
\frac{1}{m^2c^2}\left(\frac{g^{0i}p_i}{g^{00}}\right)^2
=
\frac{1}{m^2c^2}\left(\sum_{i=1}^3\frac{g_{0i}p_i}{g_{ii}}\right)^2
=o(g_{0i}).
}
Hence, the second term in \req{Htotc1} is calculated to be
\alg{
\sqrt{\left(\frac{g^{0i}p_i}{g^{00}}\right)^2-\frac{p_ip^i+m^2c^2}{g^{00}}}
&=
mc\sqrt{\left(1+\frac{2\Phi}{c^2}\right)\left(1+\frac{p_ip^i}{m^2c^2}\right)+\frac{1}{m^2c^2}\left(\frac{g^{0i}p_i}{g^{00}}\right)^2}
\\
&=
mc\left(1+\frac{\Phi}{c^2}+\frac{p_ip^i}{2m^2c^2}\right)+o\left(\frac{\Phi}{c^2},\frac{p_ip^i}{m^2c^2},g_{0i}\right)
\\
&=
mc\left(1+\frac{\Phi}{c^2}\right)+\frac{1}{2mc}\sum_{i=1}^3\frac{p_i^2}{g_{ii}}+o\left(\frac{\Phi}{c^2},\frac{p_ip^i}{m^2c^2},g_{0i}\right).
}
The approximate total Hamiltonian is thus given by
\alg{
H_{tot}({\bm q},{\bm p},E_{rest})
=
E_{rest}\left(1+\frac{\Phi({\bm q})}{c^2}\right)+\sum_{i=1}^3\left(\frac{c^2p_i^2}{2E_{rest}g_{ii}({\bm q})}-\frac{cg_{0i}({\bm q})p_i}{g_{ii}({\bm q})}\right)+o\left(\frac{\Phi}{c^2},\frac{p_ip^i}{m^2c^2},g_{0i}\right),
\laeq{htotqpE}
}
where ${\bm q}\equiv(q^1,q^2,q^3)$, and $E_{rest}$ is the rest energy of the particle.

\subsection{Derivation of A Quantum Hamiltonian}

The classical Hamiltonian \req{htotqpE} is then quantized by replacing the canonical variables $(p_i,q^i)$ with operators $(\hat{p}_i,\hat{q}^i)$ satisfying the canonical commutation relation $[\hat{p}_i,\hat{q}^j]=i\hbar\delta_{ij}$, and the rest energy $E_{rest}$ by the rest Hamiltonian $\hat{H}_{rest}$. 
Note that $(\hat{p}_i,\hat{q}^i)$ and $\hat{H}_{rest}$ act on different Hilbert spaces, so that $[\hat{H}_{rest},\hat{p}_i]=[\hat{H}_{rest},\hat{q}_i]=0$.
The total quantum Hamiltonian is given by
\alg{
\hat{H}_{tot}
=
\hat{H}_{rest}\left(1+\frac{\Phi(\hat{\bm q})}{c^2}\right)+\sum_{i=1}^3\left(\frac{c^2\hat{p}_i^2}{2\hat{H}_{rest}g_{ii}(\hat{\bm q})}-\frac{cg_{0i}(\hat{\bm q})\hat{p}_i}{g_{ii}(\hat{\bm q})}\right)_{\!\!W}.
\laeq{Htot}
}
For the ordering of operators $\{\hat{p}_i\}$ and $\{\hat{q}^i\}$ in the second term above, we adopt the {\it Weyl ordering}, as indicated by the subscript $W$ (see e.g.~Chapter 9.1 of \cite{peskin2018introduction} or Section 6 of \cite{srednicki2007quantum}; see also Section 1.2 in \cite{kashiwa1997path} for more details). 
In the Weyl ordering, each of the monomials of the canonically conjugate variables in the classical Hamiltonian is replaced by a quantum operator according to the following correspondence:
\alg{
(q^i)^m(p_i)^n
\Rightarrow
\left((\hat{q}^i)^m(\hat{p}_i)^n\right)_W
=
\left.\left(i\hbar\frac{\partial}{\partial \iota}\right)^{\! n}
\left[\exp{\left(\frac{\iota\hat{p}_i}{2i\hbar}\right)}(\hat{q}^i)^m\exp{\left(\frac{\iota\hat{p}_i}{2i\hbar}\right)}\right]\right|_{\iota=0}.
}
Noting that $\ket{q^i\pm\frac{\lambda}{2}}=\exp\left(\mp\frac{\lambda\hat{p}_i}{2i\hbar}\right)\ket{q^i}$, it can be shown that this is equivalent to
\alg{
\left((\hat{q}^i)^m(\hat{p}_i)^n\right)_W
=
\frac{1}{2\pi\hbar}\iiint dq^idp_id\lambda \: e^{\frac{ip\lambda}{\hbar}}(q^i)^m(p_i)^n\outpro{q^i+\frac{\lambda}{2}}{q^i-\frac{\lambda}{2}}.
}
Accordingly, the total quantum Hamiltonian \req{Htot} can be represented as
\alg{
\hat{H}_{tot}
=
\frac{1}{(2\pi\hbar)^3}\iiint d{\bm q}d{\bm p}d{\bm\lambda}\: \exp{\left(-\frac{{\bm p}\!\cdot\!{\bm\lambda}}{i\hbar}\right)}H_{tot}({\bm q},{\bm p},\hat{H}_{rest})\outpro{{\bm q}+\frac{\bm\lambda}{2}}{{\bm q}-\frac{\bm\lambda}{2}},
\laeq{qLam}
}
where ${\bm p}\equiv(p_1,p_2,p_3)$ and ${\bm \lambda}\equiv(\lambda^1,\lambda^2,\lambda^3)$.
Given that the eigen decomposition of the rest Hamiltonian is given by $\hat{H}_{rest}=\sum_\alpha E_\alpha\proj{\alpha}$, it is straightforward that
\alg{
\bra{\alpha',{\bm q}'}\hat{H}_{tot}\ket{\alpha,{\bm q}}
=
\frac{\delta_{\alpha',\alpha }}{(2\pi\hbar)^3}\int d{\bm p}
\exp\left(-\frac{{\bm p}\!\cdot\!({\bm q}'-{\bm q})}{i\hbar}\right)
H_{tot}\left(\frac{{\bm q}'+{\bm q}}{2},{\bm p},E_\alpha\right).
\laeq{A26}
}
The above quantization procedure based on the Weyl ordering ensures both the hermiticity of the total quantum Hamiltonian and its exact functional correspondence to the classical one.
It corresponds to the {\it mid-point prescription} when calculating the short-time kernel (see \req{bob} below).

\subsection{Calculation of The Short-Time Transition Amplitude}

Define $U(\Delta t)=\exp(\hat{H}_{tot}\Delta t/i\hbar)$.
Based on \req{A26}, the short-time transition amplitude is calculated as follows:
\alg{
&\bra{\alpha',{\bm q}'}U(\Delta t)\ket{\alpha ,{\bm q}}
\nn\\
&=\bra{\alpha',{\bm q}'}\left(\hat{I}+\frac{\hat{H}_{tot}\Delta t}{i\hbar}\right)\ket{\alpha ,{\bm q}}+O(\Delta t^2)
\\
&=\inpro{\alpha',{\bm q}'}{\alpha,{\bm q}}+\frac{\Delta t}{i\hbar}\bra{\alpha',{\bm q}'}\hat{H}_{tot}\ket{\alpha ,{\bm q}}+O(\Delta t^2)
\\
&=\frac{\delta_{\alpha',\alpha }}{(2\pi\hbar)^3}\int d{\bm p}
\left\{1+\frac{\Delta t}{i\hbar}H_{tot}\left(\frac{{\bm q}+{\bm q}'}{2},{\bm p},E_\alpha\right)\right\}\exp\left(-\frac{{\bm p}\!\cdot\!({\bm q}'-{\bm q})}{i\hbar}\right)
+O(\Delta t^2)
\\
&=\frac{\delta_{\alpha',\alpha }}{(2\pi\hbar)^3}\int d{\bm p} \exp\left[\frac{\Delta t}{i\hbar}H_{tot}\left(\frac{{\bm q}'+{\bm q}}{2},{\bm p},E_{\alpha }\right)\right]\exp\left(-\frac{{\bm p}\!\cdot\!({\bm q}'-{\bm q})}{i\hbar}\right)+O(\Delta t^2)
\\
&=\frac{\delta_{\alpha',\alpha }}{(2\pi\hbar)^3}\int d{\bm p} \exp\left[-\frac{\Delta t}{i\hbar}\left\{\frac{{\bm p}\!\cdot\!({\bm q}'-{\bm q})}{\Delta t}-H_{tot}\left(\frac{{\bm q}'+{\bm q}}{2},{\bm p},E_{\alpha }\right)\right\}\right]+O(\Delta t^2).
\laeq{bob}
}
Let $\bar{\bm q}\equiv\frac{{\bm q}'+{\bm q}}{2}$.
Using \req{htotqpE}, the exponent is calculated to be
\alg{
&\frac{{\bm p}({\bm q}'-{\bm q})}{\Delta t}-H_{tot}\left(\frac{{\bm q}'+{\bm q}}{2},{\bm p},E_{\alpha }\right)
\nn\\
&=
-E_{\alpha }\left(1+\frac{\Phi(\bar{\bm q})}{c^2}-\frac{v^2(\bar{\bm q},\bar{\bm q}',\Delta t)}{2c^2}+\sum_{i=1}^3\frac{g_{0i}(\bar{\bm q})(q'^i-q^i)}{c\Delta t}\right)-\sum_{i=1}^3\frac{c^2p_i'^2}{2E_{\alpha }g_{ii}(\bar{\bm q})}+o(g_{0i}),
}
where
\alg{
v^2(\bar{\bm q},\bar{\bm q}',\Delta t)=\sum_{i=1}^3g_{ii}(\bar{\bm q})\left(\frac{q'^i-q^i}{\Delta t}\right)^2,\quad
p_i'=p_i-\frac{E_{\alpha }}{c^2}\left(g_{ii}(\bar{\bm q})\frac{q'^i-q^i}{\Delta t}+cg_{0i}(\bar{\bm q})\right).
}

Suppose that $E_\alpha$ takes values in range $[E_0,E_0+\Delta E]$, where $E_0=m_0c^2$ is the rest energy of the particle in the ground internal state and $\Delta E$ is the energy gap scale of the internal Hamiltonian. Under the low-energy approximation (Condition (iv) above), it is assumed that $\Delta E/E_0\ll1$.
It is then straightforward that 
\alg{
\frac{c^2p_i'^2}{E_{\alpha }g_{ii}(\bar{\bm q})}
=
\frac{p_i'^2}{m_0g_{ii}(\bar{\bm q})}+o\left(\frac{p_ip^i}{m^2c^2},\frac{\Delta E}{E_0}\right).
}
Thus, we have
\alg{
&\frac{{\bm p}\!\cdot\!({\bm q}'-{\bm q})}{\Delta t}-H_{tot}\left(\frac{{\bm q}'+{\bm q}}{2},{\bm p},E_{\alpha }\right)
\nn\\
&
=
-E_{\alpha }\left(1+\frac{\Phi(\bar{\bm q})}{c^2}-\frac{v^2(\bar{\bm q},\bar{\bm q}',\Delta t)}{2c^2}+\sum_{i=1}^3\frac{g_{0i}(\bar{\bm q})(q'^i-q^i)}{c\Delta t}\right)-\sum_{i=1}^3\frac{p_i'^2}{2m_0g_{ii}(\bar{\bm q})}+o\left(\frac{p_ip^i}{m^2c^2},g_{0i},\frac{\Delta E}{E_0}\right).
\laeq{A35}
}
The higher-order terms in the above expression will henceforth be ignored.

Substituting \req{A35} to \req{bob}, and performing the Fresnel integral with respect to $p_i$, we obtain
\alg{
&\bra{\alpha',{\bm q}'}U(\Delta t)\ket{\alpha ,{\bm q}}
\nn\\
&=
\delta_{\alpha',\alpha }\left(\frac{m_0}{2\pi i\hbar\Delta t}\right)^\frac{3}{2}\sqrt{\gamma(\bar{\bm q})} \exp\left[\frac{E_{\alpha } \Delta t}{i\hbar}\left(1+\frac{\Phi(\bar{\bm q})}{c^2}-\frac{v^2(\bar{\bm q},\bar{\bm q}')}{2c^2}+\sum_{i=1}^3\frac{g_{0i}(\bar{\bm q})(q'^i-q^i)}{c\Delta t}\right)\right]+O(\Delta t^2),
}
where $\gamma(\bar{\bm q})=g_{11}(\bar{\bm q})g_{22}(\bar{\bm q})g_{33}(\bar{\bm q})$.
The exponent can be further simplified by noting that, under the assumption that $|q'^i-q^i|=O(\Delta t)$, 
\alg{
\Delta t\left(1+\frac{\Phi(\bar{\bm q})}{c^2}-\frac{v^2(\bar{\bm q},\bar{\bm q}')}{2c^2}+\sum_{i=1}^3\frac{g_{0i}(\bar{\bm q})(q'^i-q^i)}{c\Delta t}\right)
=
\Delta\tau+O(\Delta t^2)+o\left(\frac{v^2}{c^2},\frac{\Phi}{c^2},g_{0i}\right),
}
where $\Delta\tau$ is the proper time with respect to $\Delta t$, ${\bm q}'-{\bm q}$ and $g_{\mu\nu}(\bar{\bm q})$.
Indeed, we have $\Delta\tau=\frac{d\tau}{dt}\Delta t+O(\Delta t^2)$ and
\alg{
\frac{d\tau}{dt}
&=
\sqrt{-\frac{g_{\mu\nu}}{c^2}\frac{dx^\mu}{dt}\frac{dx^\nu}{dt}}
=
\sqrt{1+\frac{2\Phi}{c^2}-\frac{v^2}{c^2}-\frac{2g_{0i}}{c}\frac{dx^i}{dt}}
=
1-\frac{v^2}{2c^2}+\frac{\Phi}{c^2}-\sum_{i=1}^3\frac{g_{0i}}{c}\frac{dx^i}{dt}
+o\left(\frac{v^2}{c^2},\frac{\Phi}{c^2},g_{0i}\right).
\laeq{taudt}
}
Thus, we arrive at
\alg{
\bra{\alpha',{\bm q}'}U(\Delta t)\ket{\alpha ,{\bm q}}
=
\delta_{\alpha',\alpha }\left(\frac{m_0}{2\pi i\hbar\Delta t}\right)^\frac{3}{2}\sqrt{\gamma(\bar{\bm q})} \exp\left(\frac{E_{\alpha } \Delta \tau}{i\hbar}\right)+O(\Delta t^2).
\laeq{infProp}
}

\subsection{Derivation of The Propagator}

Take an arbitrary $N\in\mbb{N}$, let $({\bm q}_{\rm ini},t_{\rm ini})=({\bm q}_0,t_0)$, $({\bm q}_N,t_N)=({\bm q}_N,t_N)$, $\Delta t=(t_{\rm fin}-t_{\rm ini})/N$ and $t_{k+1}-t_k=\Delta t$.
We take the initial state and the final state of the internal degree of freedom to be eigenstates of the rest Hamiltonian, $\alpha_{\rm ini}=\alpha_0$ and $\alpha_{\rm fin}=\alpha_N$.
Then we have
\alg{
&\inpro{\alpha_{\rm fin},{\bm q}_{\rm fin};t_{\rm fin}}{\alpha_{\rm ini},{\bm q}_{\rm ini};t_{\rm ini}}
\nn\\
&=
\bra{\alpha_N,{\bm q}_N}\exp\left(\frac{\hat{H}_{tot}(t_N-t_0)}{i\hbar}\right)\ket{\alpha_0,{\bm q}_0}\\
&=
\bra{\alpha_N,{\bm q}_N}U(\Delta t)\left(\sum_{\alpha_{N-1}}\proj{\alpha_{N-1}}\otimes\int d{\bm q}_{N-1}\proj{{\bm q}_{N-1}}\right)U(\Delta t)\nn\\
&\quad\quad\cdots U(\Delta t)\left(\sum_{\alpha_{1}}\proj{\alpha_{1}}\otimes\int d{\bm q}_{1}\proj{{\bm q}_{1}}\right)U(\Delta t)\ket{\alpha_0,{\bm q}_0}\\
&=\sum_{\alpha_{N-1}}\cdots\sum_{\alpha_{1}}\int d{\bm q}_{N-1}\cdots\int d{\bm q}_{1} \bra{\alpha_N,{\bm q}_N}U(\Delta t)\ket{\alpha_{N-1},{\bm q}_{N-1}}\cdots\bra{\alpha_1,{\bm q}_1}U(\Delta t)\ket{\alpha_0,{\bm q}_0}\\
&=\left(\prod_{k=1}^{N-1}\sum_{\alpha_k}\int d{\bm q}_k\right)\prod_{l=0}^{N-1}\bra{\alpha_{l+1},{\bm q}_{l+1}}U(\Delta t)\ket{\alpha_l,{\bm q}_l}.
\laeq{abado}
}
Substituting this to \req{infProp} to the above, we obtain
\alg{
\inpro{\alpha_{\rm fin},{\bm q}_{\rm fin};t_{\rm fin}}{\alpha_{\rm ini},{\bm q}_{\rm ini};t_{\rm ini}}
&=\left(\prod_{k=1}^{N-1}\sum_{\alpha_k}\int d{\bm q}_k\right)\prod_{l=0}^{N-1}\left[\delta_{\alpha_{l+1},\alpha_l}\left(\frac{m_0}{2\pi i\hbar\Delta t}\right)^\frac{3}{2}\sqrt{\gamma(\bar{\bm q}_l)} \exp\left(\frac{E_{\alpha_l} \Delta\tau_l}{i\hbar}\right)\right]
\\
&=\delta_{\alpha_{\rm fin},\alpha_{\rm ini}}\left(\frac{m_0}{2\pi i\hbar\Delta t}\right)^\frac{3N}{2}\left(\prod_{k=1}^{N-1}\int d{\bm q}_k\sqrt{\gamma(\bar{\bm q}_k)}\right)\exp\left(\frac{E_{\alpha_{\rm ini}}}{i\hbar}\sum_{l=0}^{N-1}\Delta\tau_l\right),
}
where $\bar{\bm q}_l=\frac{{\bm q}_l+{\bm q}_{l+1}}{2}$.
Taking the limit of $N\rightarrow\infty$ and $\Delta t\rightarrow0$,
we arrive at
\alg{
\inpro{\alpha_{\rm fin},{\bm q}_{\rm fin};t_{\rm fin}}{\alpha_{\rm ini},{\bm q}_{\rm ini};t_{\rm ini}}
=\delta_{\alpha_{\rm fin},\alpha_{\rm ini}}\int\mathscr{D}q\sqrt{\gamma(q)}\exp\left(\frac{E_{\alpha_{\rm ini}}}{i\hbar}\int_{\rm ini}^{\rm fin}d\tau\right),
\laeq{dvorak}
}
where $\int\mathscr{D}q\sqrt{\gamma(q)}$ should be interpreted as an abbreviation of
\alg{
\lim_{N\rightarrow\infty}\left(\frac{m_0}{2\pi i\hbar\Delta t}\right)^\frac{3N}{2}\left(\prod_{k=1}^{N-1}\int d{\bm q}_k\sqrt{\gamma(\bar{\bm q}_k)}\right).
}
The inclusion of the metric factor $\sqrt{\gamma(q)}$ is necessary in order to ensure that the volume element in the integral is invariant under spatial coordinate transformations.
Finally, for arbitrary initial state $\ket{\xi_{\rm ini}}$ and $\ket{\xi_{\rm fin}}$, the propagator takes the form
\alg{
\inpro{\xi_{\rm fin},{\bm q}_{\rm fin};t_{\rm fin}}{\xi_{\rm ini},{\bm q}_{\rm ini};t_{\rm ini}}
&=\sum_{\alpha,\alpha'}\inpro{\xi_{\rm fin}}{\alpha'}\inpro{\alpha',{\bm q}_{\rm fin};t_{\rm fin}}{\alpha,{\bm q}_{\rm ini};t_{\rm ini}}\inpro{\alpha}{\xi_{\rm ini}}\\
&=
\sum_{\alpha,\alpha'}\inpro{\xi_{\rm fin}}{\alpha'}\delta_{\alpha',\alpha}\left[\int\mathscr{D}q\sqrt{\gamma(q)}\exp\left(\frac{E_{\alpha}}{i\hbar}\int_{\rm ini}^{\rm fin}d\tau\right)\right]\inpro{\alpha}{\xi_{\rm ini}}\\
&=\bra{\xi_{\rm fin}}\left[\int\mathscr{D}q\sqrt{\gamma(q)}\exp\left(\frac{\hat{H}_{rest}}{i\hbar}\int_{\rm ini}^{\rm fin}d\tau\right)\right]\ket{\xi_{\rm ini}}.
}

\section{Proof of Eqs.~\req{visQEP1}-\req{visQEP5}}
\lapp{prfvisQEP}

Noting that $|\chi_0\rangle=\cos\theta|g'\rangle+e^{i\varphi}\sin\theta|e'\rangle$, $|\chi_{1,2}\rangle=U(P_{1,2})|\chi_0\rangle$ and
\alg{
U(P_1)^\dagger U(P_2)=\exp\left(\frac{E_{g}'\Delta\tau}{i\hbar}\right)\proj{g'}+\exp\left(\frac{E_e'\Delta\tau}{i\hbar}\right)\proj{e'},
}
it is straightforward that
\alg{
\inpro{\chi_1}{\chi_2}=\bra{\chi_0}U(P_1)^\dagger U(P_2)\ket{\chi_0}=\cos^2\theta\exp\left(\frac{E_{g}'\Delta\tau}{i\hbar}\right)+\sin^2\theta\exp\left(\frac{E_e'\Delta\tau}{i\hbar}\right).
}
A further calculation yields
\alg{
\inpro{\chi_1}{\chi_2}&=\exp\left(\frac{\bar{E}'\Delta\tau}{i\hbar}\right)\left\{\cos^2\theta\exp\left(-\frac{\Delta E'\Delta\tau}{i\hbar}\right)+\sin^2\theta\exp\left(\frac{\Delta E'\Delta\tau}{i\hbar}\right)\right\}\\
&=\left\{\cos\left(\frac{\bar{E}'\Delta\tau}{\hbar}\right)-i\sin\left(\frac{\bar{E}'\Delta\tau}{\hbar}\right)\right\}\left\{\frac{1+\cos2\theta}{2}\exp\left(-\frac{\Delta E'\Delta\tau}{i\hbar}\right)+\frac{1+\cos2\theta}{2}\exp\left(\frac{\Delta E'\Delta\tau}{i\hbar}\right)\right\}\\
&=\left\{\cos\left(\frac{\bar{E}'\Delta\tau}{\hbar}\right)-i\sin\left(\frac{\bar{E}'\Delta\tau}{\hbar}\right)\right\}\left\{\cos\left(\frac{\Delta E'\Delta\tau}{\hbar}\right)+i\cos2\theta\sin\left(\frac{\Delta E'\Delta\tau}{\hbar}\right)\right\}.
}
The real part and the imaginary part of the above are calculated as
\alg{
{\rm Re}\inpro{\chi_1}{\chi_2}&=\cos\left(\frac{\Delta E'\Delta\tau}{\hbar}\right)\cos\left(\frac{\bar{E}'\Delta\tau}{\hbar}\right)+\cos2\theta\sin\left(\frac{\Delta E'\Delta\tau}{\hbar}\right)\sin\left(\frac{\bar{E}'\Delta\tau}{\hbar}\right),\\
{\rm Im}\inpro{\chi_1}{\chi_2}&=-\cos\left(\frac{\Delta E'\Delta\tau}{\hbar}\right)\sin\left(\frac{\bar{E}'\Delta\tau}{\hbar}\right)+\cos2\theta\sin\left(\frac{\Delta E'\Delta\tau}{\hbar}\right)\cos\left(\frac{\bar{E}'\Delta\tau}{\hbar}\right).
}
Let $\xi\in{\mbb R}$ be such that
\alg{
\tan{\left(\frac{\xi\Delta\tau}{\hbar}\right)}=-\cos2\theta\tan\left(\frac{\Delta E'\Delta\tau}{\hbar}\right).
}
Then we have
\alg{
\cos\left(\frac{(\bar{E}'+\xi)\Delta\tau}{\hbar}\right)
&=
\cos\left(\frac{\xi\Delta\tau}{\hbar}\right)\cos\left(\frac{\bar{E}'\Delta\tau}{\hbar}\right)-\sin\left(\frac{\xi\Delta\tau}{\hbar}\right)\sin\left(\frac{\bar{E}'\Delta\tau}{\hbar}\right)\\
&=\frac{\cos\left(\frac{\xi\Delta\tau}{\hbar}\right)}{\cos\left(\frac{\Delta E'\Delta\tau}{\hbar}\right)}\left\{\cos\left(\frac{\Delta E'\Delta\tau}{\hbar}\right)\cos\left(\frac{\bar{E}'\Delta\tau}{\hbar}\right)+\cos2\theta\sin\left(\frac{\Delta E'\Delta\tau}{\hbar}\right)\sin\left(\frac{\bar{E}'\Delta\tau}{\hbar}\right)\right\}
}
and
\alg{
\sin\left(\frac{(\bar{E}'+\xi)\Delta\tau}{\hbar}\right)
&=
\cos\left(\frac{\xi\Delta\tau}{\hbar}\right)\sin\left(\frac{\bar{E}'\Delta\tau}{\hbar}\right)+\sin\left(\frac{\xi\Delta\tau}{\hbar}\right)\cos\left(\frac{\bar{E}'\Delta\tau}{\hbar}\right)\\
&=\frac{\cos\left(\frac{\xi\Delta\tau}{\hbar}\right)}{\cos\left(\frac{\Delta E'\Delta\tau}{\hbar}\right)}\left\{\cos\left(\frac{\Delta E'\Delta\tau}{\hbar}\right)\sin\left(\frac{\bar{E}'\Delta\tau}{\hbar}\right)-\cos2\theta\sin\left(\frac{\Delta E'\Delta\tau}{\hbar}\right)\cos\left(\frac{\bar{E}'\Delta\tau}{\hbar}\right)\right\}.
}
Furthermore, we have
\alg{
\frac{\cos\left(\frac{\Delta E'\Delta\tau}{\hbar}\right)}{\cos\left(\frac{\xi\Delta\tau}{\hbar}\right)}
&=
\cos\left(\frac{\Delta E'\Delta\tau}{\hbar}\right)\sqrt{1+\tan^2\left(\frac{\xi\Delta\tau}{\hbar}\right)}\\
&=
\cos\left(\frac{\Delta E'\Delta\tau}{\hbar}\right)\sqrt{1+\cos^22\theta\tan^2\left(\frac{\Delta E'\Delta\tau}{\hbar}\right)}\\
&=
\sqrt{\cos^2\left(\frac{\Delta E'\Delta\tau}{\hbar}\right)+(1-\sin^22\theta)\sin^2\left(\frac{\Delta E'\Delta\tau}{\hbar}\right)}\\
&=
\sqrt{1-\sin^22\theta\sin^2\left(\frac{\Delta E'\Delta\tau}{\hbar}\right)}
}
Thus, letting
\alg{
\mathcal{V}_{\Delta E',\Delta\tau,\theta}
=
\sqrt{1-\sin^22\theta\sin^2\left(\frac{\Delta E'\Delta\tau}{\hbar}\right)},
}
we obtain
\alg{
{\rm Re}\inpro{\chi_1}{\chi_2}&=\mathcal{V}_{\Delta E',\Delta\tau,\theta}\cos\left(\frac{(\bar{E}'+\xi)\Delta\tau}{\hbar}\right),\\
{\rm Im}\inpro{\chi_1}{\chi_2}&=-\mathcal{V}_{\Delta E',\Delta\tau,\theta}\sin\left(\frac{(\bar{E}'+\xi)\Delta\tau}{\hbar}\right).
}
Finally, from $|\zeta_\pm\rangle=(|\chi_1\rangle\pm|\chi_2\rangle)/\sqrt{2}$,
we arrive at
\alg{
\inpro{\zeta_\pm}{\zeta_\pm}&=1\pm{\rm Re}\inpro{\chi_1}{\chi_2}=1\pm\mathcal{V}_{\Delta E',\Delta\tau,\theta}\cos\left(\frac{(\bar{E}'+\xi)\Delta\tau}{\hbar}\right),\\
\inpro{\zeta_\pm}{\zeta_\mp}&=\mp{\rm Im}\inpro{\chi_1}{\chi_2}=\pm i\mathcal{V}_{\Delta E',\Delta\tau,\theta}\sin\left(\frac{(\bar{E}'+\xi)\Delta\tau}{\hbar}\right).\laeq{visQEP3}
}

\end{widetext}

\end{document}